%% file: Project.tex
\documentclass{svjour3}

\usepackage[final]{graphicx}
\usepackage{times}
\usepackage{amsmath}
\usepackage{amssymb}
\usepackage{mathtools}
\usepackage{graphicx} 
\usepackage{epsfig}
\usepackage{ifthen}
\usepackage[utf8]{inputenc}
\usepackage[english,russian]{babel}
\usepackage[T2A]{fontenc}
\usepackage{setspace}
\usepackage{url}
\usepackage[process=auto,crossref=false]{pstool}
\usepackage[colorlinks,allcolors=blue,unicode,pdftitle={Unsteady ballistic heat transport in a 1D harmonic crystal due to a
source on an isotopic defect}]{hyperref}
\usepackage{stackrel}
\usepackage{afterpage}
\usepackage{authblk}
\usepackage[misc]{ifsym}
\usepackage{cite}

\journalname{CMAT}
\input def

\makeatletter

\@addtoreset{equation}{section}
\makeatother

\begin{document}
\selectlanguage{english}

\title{Unsteady ballistic heat transport in a 1D harmonic crystal due to a
source on an isotopic defect\thanks{This work is supported by Russian Science Support Foundation (project
21-11-00378).}}

\author{Ekaterina V. Shishkina
\and Serge N. Gavrilov}
\institute{
Ekaterina~V.~Shishkina, Serge~N.~Gavrilov (\Letter) \at
Institute for Problems in Mechanical Engineering RAS, V.O., Bolshoy pr.~61,
St.~Petersburg, 199178, Russia 
\and
Ekaterina~V.~Shishkina \at \email{shishkina\_k@mail.ru}
\and
Serge~N.~Gavrilov \at \email{serge@pdmi.ras.ru}}
%


\maketitle

\begin{abstract}
In the paper we apply asymptotic technique based on the method of
stationary phase and obtain the approximate analytical description of thermal motions 
caused by a source on an isotopic defect of an arbitrary mass in a 1D harmonic crystal. 
It is well known that localized oscillation is possible in this system in the case
of a light defect. We consider
the unsteady heat propagation 
and obtain formulae,
which provide continualization (everywhere excepting a neighbourhood of a defect)
and asymptotic uncoupling 
of the thermal motion into the sum of the slow and fast components.
The slow motion is related with ballistic heat transport, whereas the fast motion
is energy oscillation related  with transformation of the kinetic energy into
the potential one and in the opposite direction. 
To obtain the propagating
component of the fast and slow motions we estimate the exact solution in the
integral form at a moving point of observation. We demonstrate that the
propagating parts of the slow
and the fast motions are ``anti-localized'' near the defect. The physical
meaning of the anti-localization is a tendency for the unsteady propagating
wave-field to avoid a neighbourhood of a defect. 
The effect of anti-localization
increases with the absolute value of the difference between the alternated mass 
and the mass of a regular particle, and, therefore, more
energy concentrates just behind the leading wave-front of the propagating component.
The obtained solution is valid in a wide range of a spatial co-ordinate
(i.e., a particle number), everywhere excepting a neighbourhood of the leading
wave-front.
\end{abstract}


\section{Introduction}

A one-dimensional harmonic crystal (a uniform chain of particles connected by
linear springs) is a rather old classical mechanical model. Hamilton
seems to be the first who investigated
the dynamics of a uniform chain \cite{Hamilton1940}. Later this problem was
rediscovered by
Havelock \cite{Havelock1910} and Schr\"odinger \cite{schrodinger1914dynamik,Muehlich2020}.
Schr\"odinger obtained the general solution for the dynamical problem and
suggested applying this model to investigate heat transfer in crystals. The
idea was realized by many researchers, among the first works we underline
the studies by Klein \& Prigogine \cite{klein1953mecanique},
Hemmer \cite{hemmer1959dynamic},  Rieder, Lebowitz, and Lieb
\cite{rieder1967properties}. It was
shown \cite{rieder1967properties,lepri2003thermal} that heat transport in harmonic crystals
violates the Fourier low.
Nowadays, this regime of heat propagation is known as the ballistic one, and
it is experimentally observed in ultra-pure low-dimensional nano-materials
\cite{chang2008breakdown,hsiao2015micron,hsiao2013observation} under certain
conditions, in particular, in graphene 
\cite{Bae2013,Saito2018,Xu2014}. These new
applications increased again the interest to the model of a harmonic crystal.
Krivtsov, Kuzkin and their colleagues who considered the heat propagation 
in harmonic crystals
suggested separating the slow and the fast thermal processes
\cite{krivtsov2014energy,krivtsov2015heat,Kuzkin-Krivtsov-accepted,krivtsov-da70},
and formulated the simplified continuum equations involving only the slow motion.
The slow
motion is related with heat transport \cite{krivtsov2015heat,Gavrilov2019cmat},
whereas the fast motion
is energy oscillation related  with transformation of the kinetic energy into
the potential one and in the opposite direction 
\cite{klein1953mecanique,krivtsov2014energy,Gavrilov2019PhysRevE,kuzkin2019thermal}.
A comparison of the continuum approach by Krivtsov with discrete approach by
Hemmer is given in \cite{Sokolov2021}.
In this way, a general technique
was suggested that allows one to derive the simplified continuum equation for
the slow motion
and analytically investigate the unsteady and
steady-state heat transport in uniform harmonic crystals of general kind,
e.g., polyatomic lattices
\cite{Kuzkin2019}. This technique was applied, in particular, to a graphene
lattice \cite{Kuzkin2019,Gavrilov2022cmat,Panchenko2022}.
It, nevertheless, is not directly applicable to spatially non-uniform systems.
Recently Gavrilov \cite{Gavrilov2022ijhmt} noticed 
that the expression for
the slow motion in a 1D uniform chain can be formally obtained
as the slow time-varying component of the large-time asymptotics for the exact
discrete solution at a moving point of observation. 
The approach involving a moving point of observation, which is known to us due
to \cite{Slepyan1972} in context of continuum problems, 
allows one to describe running waves, wave-fronts, and 
the wave-field as a whole, in comparison with the evaluation of the
corresponding asymptotics at a fixed position. The last approach can be easily
generalized to spatially non-uniform systems.

One of the first studies concerning a 1D chain with a single isotopic defect is
\cite{Montroll1955}, where a spectral problem is considered, and it is proved that a localized mode exists in such a
system in the case of a light defect. A localized (or trapped) mode is a general
phenomenon, which can be observed in discrete and continuum systems. The
extensive bibliography on localized modes in discrete and continuum mechanical systems can be found in 
\cite{Ind-book-R2E,Andrianov2012,Gavrilov2019nody,Mishuris2020}.
Teramoto \& Takeno \cite{Teramoto1960}
were
probably the first who treated a localized oscillation in the chain with the defect as a
non-stationary non-vanishing motion in time. Kashiwamura
\cite{Kashiwamura1962} obtained a rough
estimation for the particle velocity of an initially perturbed heavy defect as
$O(t^{-3/2})$, whereas in a uniform chain the corresponding asymptotics
is $O(t^{-1/2})$, $t\to\infty$ is the time. 
Hemmer \cite{hemmer1959dynamic}, 
Magalinskii
\cite{magalinskii1959dynamical}, M\"uller 
\cite{Mueller1962} (see also \cite{Mueller2012})
considered the particular case of a very heavy isotope defect. {Turner
\cite{Turner1960} obtained the solution (the momentum autocorrelation
function) describing the motion for the defect in the case of a source on a
heavy defect as a series with terms involving Bessel functions.} For the
defect of the double 
mass\footnote{With respect to the mass of a regular particle.} 
this solution becomes a lot simpler.  Rubin published a series of studies
\cite{Rubin1960,Rubin1961,Rubin_1963}.
In \cite{Rubin1960,Rubin1961} the general problem statement for a cubic
lattice was discussed and some particular results for a 1D chain were obtained.
One-dimensional case was discussed in detail in \cite{Rubin_1963}.
It was demonstrated that
together with the localized component, which exists only for a light defect, the
solution (the momentum autocorrelation function)
always contains a vanishing propagating component. This component, which is
$O(t^{-3/2})$ at the defect (as was obtained by Kashiwamura \cite{Kashiwamura1962} for the case of a
heavy defect),
%
was calculated in the explicit form for any defect mass. Outside the
defect some particular limiting results were obtained for the case
of the defect of the double mass
(in a zone behind the leading wave-front). In the
latter case the obtained result coincides with the corresponding result for
the uniform chain (see Remark~\ref{remark-rubin}). Also, the transport of
energy was considered and the ratio of the energy trapped by the localized mode
was calculated. 
Takizawa \& Kobayasi \cite{Takizawa1967,Takizawa1968} obtained the general solution for the problem concerning
a chain with an isotope defect of an arbitrary mass as a series with
terms involving Bessel
functions. This solution has enough complicated form and is difficult for
analysis.

In \cite{Lee1989} the dynamics of a harmonic chain with an isotopic defect is considered basing on the recurrence relations
method. This method allows one to consider also diatomic chains with an
isotopic defect \cite{Yu2014,Yu2015,Yu2016}. In \cite{Yu2019} 
a chain with an impurity in mass and Hooke constant is considered by the
recurrence relations. In all these studies the solution is obtained only for
the defect particle.

In several studies
\cite{kannan2013heat,Paul2020,Gendelman2021,Plyukhin2020} the classical
steady-state formulation of the problem concerning thermal conductivity of a
finite chain with an isotopic defect is considered. Gendelman
\& Paul \cite{Paul2020,Gendelman2021} suggested using this model to describe
the Kapitza thermal resistance.  Plyukhin \cite{Plyukhin2020} indicated that
in the framework of the model non-Clausius heat transfer (from cold to hot) is
possible. Experimental studies (e.g., \cite{Chen2012}) demonstrate that the presence of
isotopic defects essentially influences on the thermal conductivity of a pure
graphene. Phonon scattering by an isotope defect in graphene is considered in
\cite{Saito2018}.

In this paper we apply the asymptotic technique based on the method of stationary
phase to an integral representation for the solution and obtain the
approximate analytical description of the thermal motion caused by a source on
an isotopic defect in a 1D harmonic crystal. We re-obtain some necessary known
results, mostly related with localized oscillation and asymptotics of the
solution just on a defect, to get necessary intermediate formulae and make the
analysis more complete. 

The new results of the paper are related mostly with the
propagating component of the motion. These results have been obtained due to a
mathematical approach, which is well-known for continuum systems \cite{Slepyan1972}, 
but, as far as we know, seems to be quite new \cite{Gavrilov2022ijhmt}
for discrete systems. Following to
\cite{Gavrilov2022ijhmt}, we estimate the exact solution in the integral form
at a moving point of observation. For a discrete system, the new approach directly provides
continualization and slow-and-fast motions decoupling, as well as a possiblity 
to obtain the approximate solution in a wide range of a spatial co-ordinate (i.e., a
particle number), whereas the previous solutions describe only the isotopic
defect motion and leading wave-fronts. The obtained solution has a simple structure and is valid for
any value of the defect mass, although it, as well as all
previous solutions, 
are not applicable in a neighbourhood of the defect with the mass close to the mass
of a regular particle.
It demonstrates us that the
propagating parts of the slow
and the fast motions are ``anti-localized'' near the defect. The physical
meaning of the anti-localization is a tendency for the unsteady propagating
wave-field to avoid a neighbourhood of a defect. 
The effect of anti-localization
increases with the absolute value of the difference between the alternated mass 
and the mass of a regular particle, and, therefore, more
energy concentrates just behind the leading wave-front of the propagating component. 
For the best of our knowledge, this important qualitative finding is also new.
The separation of slow motions allows
one to essentially simplify 
\cite{krivtsov2015heat,Kuzkin-Krivtsov-accepted,krivtsov-da70,Gavrilov2019cmat,Sokolov2021,Kuzkin2019,Gavrilov2022cmat}
the description of the ballistic heat transfer 
and gives the possibility to consider complicated systems, in particular,
2D--3D polyatomic lattices. 
Therefore, we expect that our approach can also be
applied to such a system with defects. 

Thus, two ideas that allow us to
obtain the more complete and simple description of ballistic heat transfer in the system
under consideration are
\begin{itemize} 
\item Separation of slow motions, suggested in \cite{krivtsov2015heat};
\item Estimation of the exact solution in the integral form at a continuously
moving point of observation, suggested for a discrete system in
\cite{Gavrilov2022ijhmt}.
\end{itemize}
It is important that both of the ideas are applicable for more complicated
lattices with defects.

The paper is organized as follows. In Sect.~\ref{sec-notation} we introduce
general notation. In Sect.~\ref{sec-formulation} we discuss the mathematical
formulation. In Sect.~\ref{sec-formulation1} we introduce an auxiliary
deterministic problem for the chain with an isotopic defect,
and the corresponding fundamental solution describing the particle velocity. 
The momentum autocorrelation function, which is traditionally used to describe
the motion in the system, equals this fundamental solution
with accuracy to a constant multiplier.
In Sect.~\ref{sec-formulation2} we formulate the thermal (random) problem and
introduce the corresponding fundamental solution, 
describing the propagation of the kinetic
temperature along the chain. This thermal fundamental solution can be
expressed in terms of the squared fundamental solution describing the particle velocity.
In Sect.~\ref{sec-disp} we discuss the dispersion relation and its properties
in the pass-band and the stop-band.
In Sect.~\ref{App-with} we introduce the Green function in the frequency
domain for the pass-band and the stop-band. In Sect.~\ref{sec-spectral-p}
we consider the spectral problem for the localized mode and get the formula
for the frequency of localized oscillation. The material in
Sects.~\ref{sec-disp}--\ref{sec-spectral-p} is definitely not new. In
Sect.~\ref{sec-non} we get the integral representation for the fundamental
solution of the non-stationary deterministic problem for the particle velocity
and proceed with asymptotic evaluation. 
In Sect.~\ref{sect-pass} we calculate the contribution from the pass-band (the
propagating vanishing component) using the stationary phase method applied on
a moving point of observation. These results are new. 
In Sect.~\ref{sband-co}  we calculate the contribution from the stop-band (a
localized oscillation). The final results of Sect.~\ref{sband-co} are not new, 
though we use the mathematical technique a bit different 
from the traditional one. Namely, we prefer to use 
asymptotic integration over
the real line, using the theory of Fourier transform for generalized functions (or
distributions), the limit absorption principle and the Sokhotski–Plemelj
theorem for the real line, instead of Laplace transform and 
integration over a
contour in the complex plane. In Sect.~\ref{bou-contrib} we calculate the contribution 
from the cut-off frequency and re-obtain the asymptotics for the particle
velocity of the defect got by Rubin \cite{Rubin_1963}.
In Sect.~\ref{sec-onfront} we calculate the corresponding asymptotics on the
leading wave-front.
In Sect.~\ref{n>t} we discuss the asymptotics before the leading wave-fronts,
this result seems to be new.
To make the results 
reproducible, we
present detailed calculations for all asymptotics. For a reader
who prefers to skip the details, it is possible to find the most important formulae
summarized in Sect.~\ref{sect-before-front}. 
In Sect.~\ref{sec-num} we compare the obtained results with numerical ones
and demonstrate a good agreement. In Sect.~\ref{sect-disc} we discuss the
domain of applicability for our results and compare them with ones
previously obtained in \cite{Rubin_1963}.
In Sect.~\ref{sec-thermal} we {return to} the thermal problem and obtain formulae,
which provide continualization and asymptotic uncoupling 
of the thermal motion into the sum of the slow and fast components. In 
Sect.~\ref{sec-conc} we discuss the basic results of the paper and their possible
generalizations. 
In Appendix~\ref{app-nondim} we derive non-dimensional formulation, which is
used everywhere in the paper.
In Appendix~\ref{sec-erdelyi} we provide the formulation of
the Erd\'elyi lemma, which is used to calculate the asymptotics. In 
Appendix~\ref{sec-ratio} we calculate the trapped energy ratio. This result is
previously obtained in~\cite{Rubin_1963}, but seems to be very important for
understanding. 

\section{Nomenclature}
\label{sec-notation}
In the paper, we use the following general notation:
\begin{description}	
\item[$\mathbb Z$] is
the set of all integers;
\item[$\mathbb R$] is
the set of all real numbers; 
\item[$C^\infty$] is the set of all infinitely differentiable functions;
\item[$C_0^\infty$] is the set of all finite infinitely differentiable functions;
\item[$H(\cdot)$] is the Heaviside function;
\item[$\langle\cdot\rangle$] is the mathematical expectation for a random quantity;
\item[$\de_n$ ] is {the Kronecker delta} ($1$ if only if $n=0$, $0$ otherwise, {$n\in\mathbb Z$});
\item[$\de(\cdot)$ ] is {the Dirac delta-function}; 
\item[$k_B$] is the (dimensionless) Boltzmann constant;
\item[$J_n(\cdot)$] is the Bessel function of the first kind of order $n$;
\item[$\Gamma(\cdot)$] is the Gamma function;
\item[$\PV$] is the Cauchy principal value for an integral;
\item[$\cc$] are the complex conjugate terms;
\item[$m$] is the dimensionless mass of an isotopic defect.

\end{description}	

\section{The problem formulation}
\label{sec-formulation}
\subsection{The deterministic problem}
\label{sec-formulation1}
Consider a chain of mass points of an equal mass with one alternated mass. 
All masses are connected by linear springs with the same stiffness.
The equations of motions in the dimensionless form can be expressed as the following 
infinite system of differential-difference equations:
\begin{gather}
m_n\ddot{u}_n-(u_{n+1}-2u_n+u_{n-1})= \delta_n p(t).
\label{chain-eq-basic-al0}
\end{gather}
Here 
$n \in \mathbb{Z}$,
$u_n(t)$ is the dimensionless displacement of the particle with a number
$n$,
$m_n$ is
the dimensionless mass of the particle with number $n$:
\begin{gather}
m_n=1+\delta_n(m-1),
\label{m_n}
\end{gather}
overdot denotes the derivative with respect to
the dimensionless time $t$. We assume that the dimensionless mass of the defect
particle $m=m_0$ is such that
\begin{gather}
m>0, \qquad m\neq1.
\end{gather}
The dimensionless external force $p(t)$ is applied to  
the zeroth particle
with alternated mass.
Equation~\eqref{chain-eq-basic-al0}
can be transformed to the following equivalent form:
\begin{gather}
\ddot{u}_n-(u_{n+1}-2u_{n}+u_{n-1})=\delta_n P(t),
\label{chain-gov-eq-loaded}
\\
P(t)=-(\DM) \ddot{u}_0+p(t),
\label{external-load-p(t)}
\end{gather}
where $P(t)$ is an unknown function.
The differential-difference operator in the left-hand side of 
Eq.~\eqref{chain-gov-eq-loaded} corresponds to a 
uniform chain of mass points of unit mass connected by springs of unit
stiffness.

\begin{remark}  
In the paper we use the dimensionless problem formulation from the very beginning.
Non-dimensionalization is discussed in Appendix~\ref{app-nondim}.
\end{remark}

Since we are interested mostly in thermal processes, which are related with
the propagation of the kinetic energy (or kinetic temperature), 
in what follows, we deal with the
expression for the particle velocity $\dot u_n$.
In the paper we use the fundamental solution 
\begin{equation}
u_n=\UF_n,
\qquad
\dot u_n=\VF_n\=\dot\UF_n
\end{equation}
of the
deterministic problem, which corresponds to the choice of the external force 
as the pulse force
\begin{equation}
p(t)=\delta(t).
\label{p-detlta-t}
\end{equation}
In this case the initial conditions for Eq.~\eqref{chain-eq-basic-al0}
can be
formulated in the following form, which is conventional for distributions (or
generalized functions) \cite{Vladimirov1971}:
\begin{equation}
u_n \big|_{t<0} \equiv 0.
\label{initial-cond}
\end{equation}

The generalized initial value problem 
\eqref{chain-eq-basic-al0} with the right-hand side 
defined by Eq.~\eqref{p-detlta-t}
and initial conditions Eq.~\eqref{initial-cond}
can be equivalently formulated in the form of classical initial value problem
for the system of equations
\begin{equation}
\begin{gathered}        
m_n\dot{\VF}_n-(\UF_{n+1}-2\UF_n+\UF_{n-1})= 0,\\
\dot \UF_n=\VF_n
\end{gathered}
\label{chain-eq-basic-class}
\end{equation}
with initial conditions in the classical form \cite{Vladimirov1971}
\begin{gather}
\UF_n(0)=0,\qquad \VF_n(0)=m_n^{-1}\delta_n\equiv m^{-1}\delta_n.
\end{gather}

In the particular case $m=1$ of a uniform chain the exact expression for the fundamental solution 
$\VF_n$ is \cite{schrodinger1914dynamik}
\begin{equation}
\VF_n=J_{2n}(2t).
\label{Sro-bessel}
\end{equation}

\subsection{The random (thermal) problem}
\label{sec-formulation2}
Consider the case of a point random initial excitation.
{Let the initial conditions for Eq.~\eqref{chain-eq-basic-class} be as
follows:}
\begin{gather}  
u_n(0)=0,\qquad \dot u_n(0)=\rho{\delta_{n}}.
\end{gather}
Here $n\in\mathbb Z$, $\rho$ is a random quantity such that 
\begin{gather}
\langle\rho\rangle=0,\qquad \langle\rho^2\rangle=\sigma.
\label{sol-cov-general}
\end{gather}
The (dimensionless) kinetic temperature $T_n$ is conventionally introduced by the
following formula:
\begin{equation}
T_n\=
2k_B^{-1}\langle K_n\rangle,
\label{sol-T-def}
\end{equation}
where 
\begin{gather}
K_n(t)=\frac{m_n\dot u_n^2(t)}2
\end{gather}
is the kinetic energy,
\begin{gather}
\langle K_n(t)\rangle=
\frac{m_n}2\,\langle \dot u^2_n \rangle 
=
\frac{\sigma m_nm^2}2\, \VF_n^2(t)
\end{gather}
is the mathematical expectation for the kinetic energy,
\begin{gather}
\sum_n\langle K_n(0)\rangle=
\frac{\sigma m^3}2\, \VF_0^2(0)
=
\frac{\sigma m}2\, 
=
\mathscr E
\label{E-def}
\end{gather}
is the mathematical expectation for the initial kinetic (as well as the total)
energy for the whole harmonic crystal.
Thus,
\begin{equation}
\langle K_n(t)\rangle=\mathscr E m m_n  \VF_n^2(t),
\end{equation}
and, therefore,
\begin{gather}
T_n(t)=
k_B^{-1} \mathscr E 
\,\TT_n(t),
\label{Tc-def}
\end{gather}
where we call the quantity
\begin{gather}
\TT_n(t)=
2m m_n \VF_n^2(t)
\label{thermal-fs}
\end{gather}
the thermal fundamental solution. One has 
\begin{equation}
\TT_n(0)=2\delta_n.
\label{2dn}
\end{equation}
We choose the fundamental solution $\TT_n$, which satisfies the initial
normalization condition \eqref{2dn}, to be in agreement with the previous studies 
\cite{krivtsov2015heat,krivtsov-da70,Gavrilov2019cmat,Sokolov2021,Gavrilov2022ijhmt},
where a uniform chain is under consideration. 
In that particular case $m=1$ the exact expression for 
the thermal fundamental solution 
$\TT_n$ is \cite{Sokolov2021,Gavrilov2022ijhmt}
\begin{equation}
\TT_n=2J^2_{2n}(2t).
\end{equation}

\begin{remark}
{In many studies}, in particular in
\cite{Kashiwamura1962,Rubin_1963,Yu2014,Yu2015,Yu2016,Yu2019}, the
particle velocity is characterized in the framework of the thermal problem. To
make this possible, the momentum autocorrelation function,
which equals the fundamental solution $\VF_n$ with accuracy to a
constant multiplier, has been introduced.
\end{remark}

\section{The dispersion relation}
\label{sec-disp}
To obtain the dispersion relation for a uniform chain
we put $m=1$, $p=0$ in 
Eqs.~\eqref{chain-eq-basic-al0}, \eqref{m_n}
and look for the solution in the following form:
\begin{equation}
u_n=U_0\EXP{-\I (\Omega t+ qn)},
\label{form-osc}
\end{equation}
where $\Omega \in \mathbb{R}$ is the frequency,
$q$ 
is the wave-number.
Substituting this expression into Eq.~\eqref{chain-eq-basic-al0}, one gets:
\begin{equation}
-\Omega^2+2=\EXP{-\I q}+\EXP{\I q}
\label{dis-form1-al0}
\end{equation}
or
\begin{equation}
\Omega^2=4\sin^2\frac{q}{2}\equiv 2(1-\cos q).
\label{dis-form3-al0}
\end{equation}
Thus, the whole frequency band $\Omega\in\mathbb R$ can be divided to the
pass-band 
\begin{equation}
\mathbb P\=[-\Omega_\ast,\Omega_\ast], 
\end{equation}
where the corresponding wave-numbers $q(\Omega)$
are reals, and the stop-band 
\begin{equation}
\mathbb S\=(-\infty,-\Omega_\ast)\cup(\Omega_\ast,\infty),
\end{equation}
where the corresponding wave-numbers are imaginary.
Here 
\begin{equation}
\Omega_\ast\=2
\end{equation}
is the cut-off (or boundary) frequency, which separates the bands.
Put also
\begin{equation}
\mathbb P_+\=[0,\Omega_\ast], 
\qquad
\mathbb S_+\=(\Omega_\ast,\infty).
\end{equation}

\begin{figure}[htb]  
\centering\includegraphics[width=0.8\textwidth]{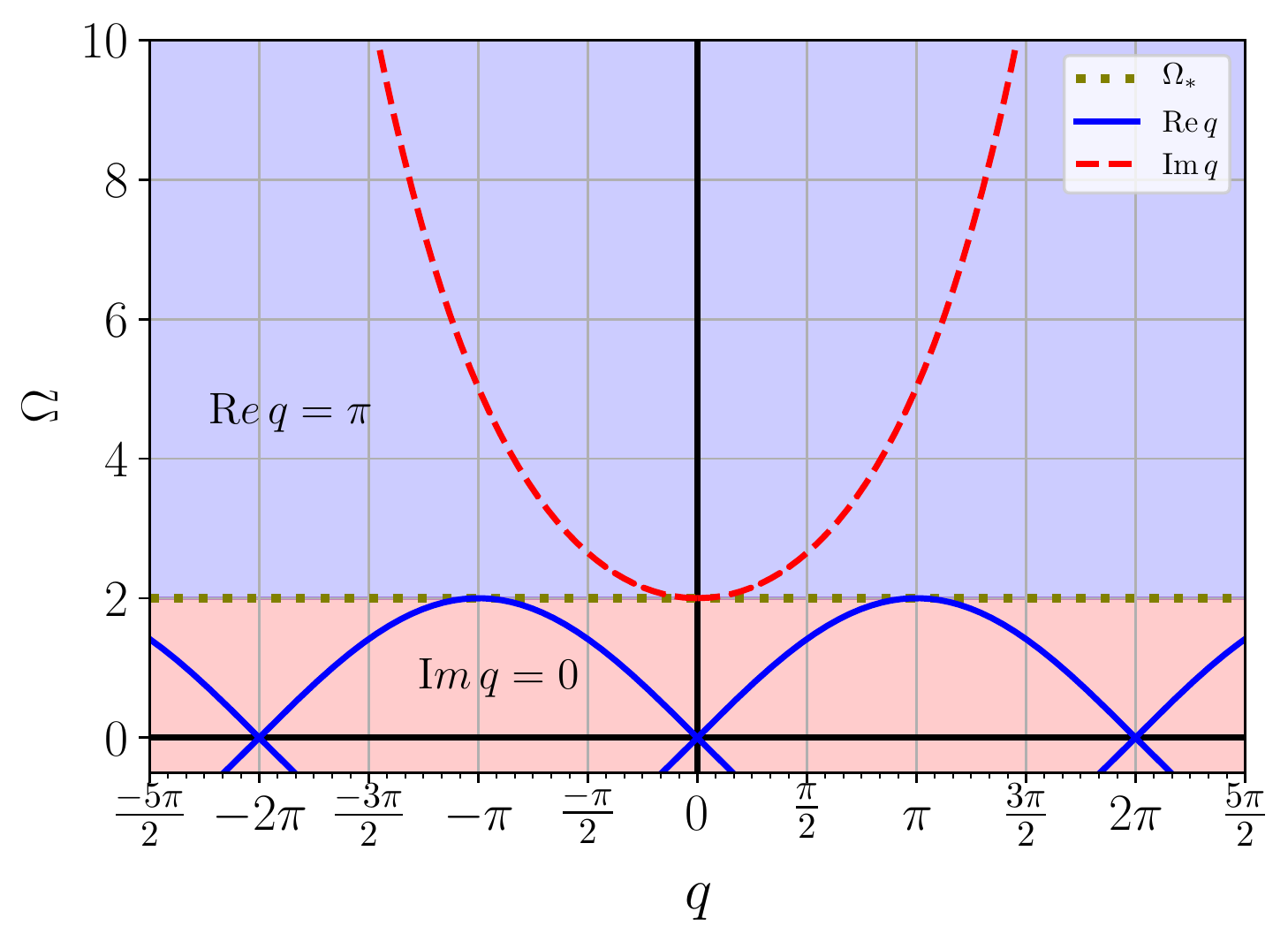}
\caption{Dispersion relation: the real and the imaginary parts of the
wave number $q$ versus frequency $\Omega$}
\label{dispersion.pdf}
\end{figure}

For $\Omega\in\mathbb P$ according to 
\eqref{dis-form3-al0}
one has 
\begin{equation}
q=\pm a,\qquad a=\arccos \frac{2-\Omega^2}2.
\label{wn-pass}
\end{equation}
%
%
For $\Omega\in\mathbb S$ 
one has $\Im q\neq0$. 
Due to Eq.~\eqref{dis-form1-al0} the
dispersion equation can be rewritten as follows:
\begin{equation}
-\Omega^2+2=2\cos (\Re q) \cosh (\Im q) -2\I\sin (\Re q) \sinh (\Im q).
\end{equation}
Since $\Omega \in \mathbb{R}$, the imagery part of the right-hand side of the
last equation should be zero:
\begin{equation}
2\sin (\Re q) \sinh (\Im q) =0.
\end{equation}
One has $\sinh (\Im q) \neq 0$, hence,
$\sin (\Re q)=0$.
Since $-\Omega^2+2<0$ for $\Omega\in\mathbb S$ {one can take} 
\begin{gather}
\Re q=\pi,
\label{a-pi}
\\
\EXP {\I n \Re q} =(-1)^n,
\label{a-value}
\end{gather}
i.e., neighbouring particles oscillate in counter phase.
Hence, the dispersion relation can be written as follows:
\begin{equation}
-\Omega^2+2=-2\cosh (\Im q).
\label{dis-complex-wn-form3-al0}
\end{equation}
Thus,
\begin{gather}
q=\pi\pm \I b,
\label{q-pi-b}
\\
b=
\arccosh\frac{1}{2}(\Omega^2-2)
=
\ln 
\left( \frac{1}{2}(\Omega^2-2)  
+\sqrt{\frac{1}{4}(\Omega^2-2)^2-1}
\right).
\label{wn-stop}
\end{gather}
In Fig.~\ref{dispersion.pdf} the plot for the real and the imaginary parts of the
wave number $q$ versus frequency $\Omega$ calculated according to the
dispersion relation 
\eqref{dis-form3-al0} in the pass-band and the stop-band
are presented.

\section{The Green function in the frequency domain}
\label{App-with}
Consider now Eqs.~\eqref{chain-gov-eq-loaded}, \eqref{external-load-p(t)},
wherein 
\begin{gather}
u_n(t) = U_n(\Omega)\, \mathrm{e}^{-\I \Omega t},
\label{u-harmonic}
\\
p(t)=\mathrm e^{-\I\Omega t}.
\label{harmonic-Uscr}
\end{gather}
We substitute expressions~\eqref{u-harmonic}, \eqref{harmonic-Uscr}
into Eqs.~\eqref{chain-gov-eq-loaded}, 
\eqref{external-load-p(t)}.  This yields
\begin{equation}
(-\Omega^2+2)U_n-U_{n+1}-U_{n-1}=\delta_n\big(\Omega^2(\DM)U_0 +1\big).
\label{chain-gov-eq-loaded-Four1}
\end{equation}
The corresponding steady-state solution $U_n=\mathscr G_n$ of the obtained equation is 
the Green function in the frequency domain for a chain with an isotopic defect.
We look for the solution in the following form:
\begin{align}
&
{U}_n=U_0(\Omega)\,\EXP{\I q |n| \sign \Omega}, \quad &&q=a(\Omega),&\quad
&\Omega\in\mathbb P;
\label{sol-pass-band-Four}
\\
&
{U}_n=U_0(\Omega)\,\EXP{\I q |n|}, \quad &&q=\pi+\I b(\Omega),&\quad 
&\Omega\in\mathbb S;
\label{u_k-amp-loaded}
\end{align}
where the wave-number $q$ 
is defined in accordance with
Eq.~\eqref{wn-pass} and
Eqs.~\eqref{q-pi-b}, \eqref{wn-stop} for the pass-band and the stop-band,
respectively.  Expression \eqref{sol-pass-band-Four} 
satisfies the Sommerfeld radiation conditions, whereas
Eq.~\eqref{u_k-amp-loaded} satisfies vanishing boundary conditions at
infinity.
For $|n|\geq1$ equations from the system \eqref{chain-gov-eq-loaded-Four1}
transform into corresponding homogeneous equations, which are clearly
satisfied by exponential functions 
\eqref{sol-pass-band-Four},
\eqref{u_k-amp-loaded}. To find unknown $U_0$ we need to consider
the equation corresponding to $n=0$.
This yields
\begin{align}
&
\mathscr G_n(\Omega)=\frac{\EXP{\I a |n|\sign \Omega}}
{-m\Omega^2-2\EXP{\I a \sign \Omega}+2}
,&\quad
&\Omega\in\mathbb P;
\label{Green-function0-lower}
\\
&
\mathscr G_n(\Omega)=\frac{(-1)^{|n|}\EXP{-b|n|}}{-m\Omega^2+2\EXP{-b}+2}
,&\quad
&\Omega\in\mathbb S.
\label{Green-function0-upper}
\end{align}
Substituting Eq.~\eqref{wn-pass} and Eq.~\eqref{wn-stop}
into Eq.~\eqref{Green-function0-lower}
and Eq.~\eqref{Green-function0-upper}, respectively, yields:
\begin{align}
&
\mathscr G_n(\Omega)=-\frac{
\EXP{\I  |n|\sign \Omega\, \arccos\frac{2-\Omega^2}2}}
{(\DM)\Omega^2+\I\Omega \sqrt{4-\Omega^2}}
,&\quad
&\Omega\in\mathbb P;
\label{Green-function0-lower-e}
\\
&
\mathscr G_n(\Omega)=\frac{(-1)^{|n|}2^{|n|}}
{\Phi^{|n|-1}(\Omega)\big((-m\Omega^2+2)\Phi(\Omega)+4\big)}
,&\quad
&\Omega\in\mathbb S;
\label{Green-function0-upper-e}
\end{align}
where
\begin{equation}
\Phi(\Omega)\=\Omega^2-2+|\Omega|\sqrt{\Omega^2-4}.
\end{equation}

In the particular case $m=1$ we have obtained the Green function in the
frequency domain for a uniform chain:
\begin{equation}
G_n(\Omega)\=\mathscr G_n(\Omega)\big|_{m=1}.
\label{G-def}
\end{equation}
Note that it is easy to see that the cut-off frequency $\Omega_\ast=2$ 
is a resonant frequency for a
a uniform chain.
{The corresponding non-stationary solution
grows as $t\to\infty$}
\cite{hemmer1959dynamic,slepyan1987energy,AyzenbergStepanenko2008,Abdukadirov_2019}.

\section{The spectral problem for the localized mode}
\label{sec-spectral-p}

Putting $p = 0$ and assuming 
that Eq.~\eqref{u-harmonic} is
fulfilled,
consider the steady-state problem concerning the natural
localized oscillation of the system described by Eqs.~\eqref{chain-gov-eq-loaded},
\eqref{external-load-p(t)} at positive\footnote{Since the
equations of motions involve only second order time derivatives.} frequency $\Omega_0$.
Since we look for a mode with finite energy we require that
\begin{equation}
\sum_{n=-\infty}^{\infty} U_n^2<\infty,\qquad
\sum_{n=-\infty}^{\infty} (U_{n+1}-U_n)^2<\infty
\end{equation}
and consider the 
frequencies in the stop-band: $\Omega_0\in\mathbb S_+$.
We substitute expression~\eqref{u-harmonic}
into Eqs.~\eqref{chain-gov-eq-loaded},
\eqref{external-load-p(t)}. This yields 
\begin{equation}
(-\Omega_0^2+2)U_n-U_{n+1}-U_{n-1}=\delta_n\Omega_0^2 (\DM)U_0.
\label{chain-gov-eq-loaded-Four-p0}
\end{equation} 
The solution of the above equation can be written as follows:
\begin{equation}
U_n=(\DM) \Omega_0^2  G_n(\Omega_0)U_0,
\label{ampl-sol}
\end{equation}
where $G_n(\Omega)$ is the Green function for a uniform chain given by
Eq.~\eqref{G-def}.
Calculating Eq.~\eqref{ampl-sol}
at $n=0$, one can derive the following frequency equation for the frequency
$\Omega_0$ of the localized mode:
\begin{equation}
\mathscr G_0^{-1}(\Omega_0)=0,
\label{freq-eq-complex-wn-al0-0}
\end{equation}
i.e., $\Omega_0$ should be a resonant frequency.
The last equation can be equivalently transformed to the following form, which
is useful for the further treatment:
\begin{equation}
\frac{1}{2}(m\Omega_0^2-2)=\EXP{-b(\Omega_0)},
\label{freq-eq-complex-wn-al0}
\end{equation}
where $b(\Omega_0)>0$ is defined by the right-hand side of 
Eq.~\eqref{wn-stop}.

To find the necessary and sufficient condition for the existence of the localized
mode, it is useful to consider simultaneously the frequency
equation~\eqref{freq-eq-complex-wn-al0} and the dispersion
equation~\eqref{dis-complex-wn-form3-al0}. These two equations can be rewritten 
in the following form:
\begin{gather}
\Omega_0^2-2=2\cosh b(\Omega_0),
\label{eq-complex-wn-1-al0}\\
(m-1)\Omega_0^2=-2\sinh b(\Omega_0).
\label{eq-complex-wn-2-al0}
\end{gather}
One can see that for $m>1$ Eq.~\eqref{eq-complex-wn-2-al0} cannot be fulfilled
since $b>0$ due to 
\eqref{u_k-amp-loaded}.
Thus, the localized mode does not exist in this case.

Consider the case $m<1$. For $\Omega_0 \in \mathbb S_+$ both
Eqs.~\eqref{eq-complex-wn-1-al0}, \eqref{eq-complex-wn-2-al0} can be fulfilled.
Squaring Eqs.~\eqref{eq-complex-wn-1-al0}, \eqref{eq-complex-wn-2-al0} and
subtracting the second squared equation from the first one, we rewrite 
the frequency equation in the following equivalent form
\begin{equation}
\Omega_0^2(\Omega_0^2m(2-m)-4)=0.
\label{freq-eq-complex-al0}
\end{equation}
The only positive root of this equation is
\begin{equation}
\Omega_0=\frac2{\sqrt{m(2-m)}}.
\label{loc-freq-m<1-zero}
\end{equation}
Now we should verify that $\Omega_0$ given by Eq.~\eqref{loc-freq-m<1-zero}
belongs to $\mathbb S_+$, i.e., $\Omega_0>\Omega_\ast$. Squaring 
Eq.~\eqref{loc-freq-m<1-zero} gets
\begin{equation}
\frac4{m(2-m)} > 4\quad\Longleftrightarrow\quad
(m-1)^2 > 0,
\end{equation}
which is true.
Thus, for all $m<1$ there exists a unique localized mode with frequency
$\Omega_0$ defined by 
Eq.~\eqref{loc-freq-m<1-zero}.

The corresponding form of localized oscillation can be found by 
Eq.~\eqref{ampl-sol}. One has
\begin{equation}
U_n=U_0(-1)^{|n|}\EXP{-b(\Omega_0)|n|},
\end{equation}
where $U_0$ is an arbitrary constant. 
Taking into account Eqs.~\eqref{freq-eq-complex-wn-al0},
\eqref{loc-freq-m<1-zero}
one gets:
\begin{equation}
U_n=U_0\left(\frac{-m}{2-m} \right)^{|n|}.
\label{loc-osc-form-m<1-zero}
\end{equation}

%

%
%

\section{Asymptotics for the fundamental solution $\mathscr V_n$ of the deterministic problem}
\label{sec-non}

{To obtain the expression for the fundamental solution $\UF_n$,
we apply the Fourier transform with respect to time $t$ to
Eqs.~\eqref{chain-gov-eq-loaded}, \eqref{external-load-p(t)}, and 
obtain in this way Eq.~\eqref{chain-gov-eq-loaded-Four1}, wherein $U_n$ has the 
meaning of the Fourier transform with respect to time~$t$ of $\UF_n$.
The solution of this equation is the Green function $\mathscr G_n$ given by 
Eqs.~\eqref{Green-function0-lower}, \eqref{Green-function0-upper}.}
%
Now $\UF_n(t)$ can be represented as the inverse Fourier transform:
\begin{equation}
\UF_n=\frac{1}{2\pi}\left(\int_{\mathbb P}+\int_{\mathbb S}\right)
\mathscr G_n(\Omega)\pFO\EXP{-\I\Omega t} \, \d\Omega=
\UF_n^\mathrm{pass}
+
\UF_n^\mathrm{stop}.
\label{u-fourier-gen}
\end{equation}
Accordingly,
\begin{multline}
\VF_n
=
\VF_n^\mathrm{pass}
+
\VF_n^\mathrm{stop}
=-\frac{\I}{2\pi}\left(\int_{\mathbb P}+\int_{\mathbb S}\right)
\Omega\mathscr G_n(\Omega)\pFO\EXP{-\I\Omega t} \, \d\Omega\\=
-\frac{\I}{2\pi}\left(\int_{\mathbb P_+}+\int_{\mathbb S_+}\right)
\Omega\mathscr G_n(\Omega)\pFO\EXP{-\I\Omega t} \, \d\Omega+\cc=
I_n^{\mathrm{pass}}+I_n^{\mathrm{stop}}+\cc
\label{dot-u-fourier-gen}
\end{multline}

\begin{remark} 
In the case $m<1$, where the localized mode exists (see Sect.~\ref{sec-spectral-p}), the integral 
\eqref{dot-u-fourier-gen}
over the stop-band does not exist in the
classical sense
due to the poles $\pm\Omega_0$ of the Green function $\mathscr G_n$ on the real axis.
The corresponding details on the meaning of the 
integral in this case are provided in Sect.~\ref{sband-co}. The integrand of
\eqref{u-fourier-gen} (but not \eqref{dot-u-fourier-gen}) also has a pole at
$\Omega=0$, which corresponds to the chain motion as a whole.
\end{remark}

The integrals 
$I_n^{\mathrm{pass}}$ and $I_n^{\mathrm{stop}}$ have the structure of a Fourier integral:
\begin{equation}
I=\int A(\Omega)\,\EXP{\I\phi(\Omega)t} \,\d\Omega.
\end{equation}
To estimate them, we use, in what follows, the procedure of asymptotic
evaluation for large times based on the method of stationary
phase \cite{erdelyi1956asymptotic,Fedoryuk1977,temme2014}. 

{\it To make the results to be
reproducible, we
present detailed calculations for all the corresponding asymptotics.
For a reader who prefers to skip the details, it is possible
to find the most important formulae summarized in Sect.~\ref{sect-before-front}.}

Asymptotics of a Fourier integral is the sum of 
contributions $I(\Omega_i)$ from the critical points $\Omega_i$:
\begin{gather}
I=\sum_i I(\Omega_i)+O(t^{-\infty}),\qquad t \to\infty;
\\
I(\Omega_i)\=
\int \mathscr A(\Omega)\chi_{\Omega_i}(\Omega)\EXP{\I\phi(\Omega)t} \,\d\Omega.
\label{contrib}
\end{gather}
The critical points are stationary points for the phase $\phi(\Omega)$, 
finite end-points of
the integration intervals and singular points for the phase $\phi(\Omega)$ and the
amplitude $\mathscr A(\Omega)$. Here $\chi_{\Omega_i}(\Omega)$ is a
neutraliser at  $\Omega=\Omega_i$ such that
$\chi_{\Omega_i}(\Omega)\equiv0$ in a neighbourhood of any $\Omega_j$
for $j\neq i$.
\label{neu-def}
\begin{remark}  
{Neutraliser \cite{temme2014,van1948method} $\chi_{\Omega_i}(\Omega)$ at a critical point $\Omega_i$ is a
$C_0^\infty(\mathbb R)$ function such that
$\chi_{\Omega_i}(\Omega_i)=1$, $\chi^{(n)}_{\Omega_i}(\Omega_i)=0$ for $n>1$,
and $\chi_{\Omega_i}(\Omega)\equiv0$ outside of a certain neighbourhood of
$\Omega_i$.}
\end{remark}

\subsection{The case $0<|n|<t$: the contribution from the pass-band}
\label{sect-pass}

One has:
\begin{gather}
\Idva_n=-\frac{\I }{2\pi}\int_{0}^{2}\pFO
\frac{\EXP{\I |n|\arccos \frac{2-\Omega^2}2-\I\Omega t}}
{-(\DM)\Omega-\I\sqrt{4-\Omega^2}}\, \d \Omega.
\label{I-2-pass-band}
\end{gather}

Since the contribution from the pass-band describes propagating waves,
following to \cite{Gavrilov2022ijhmt}, we estimate the large-time asymptotics 
of the right-hand side of 
Eq.~\eqref{I-2-pass-band}
at the moving front
\begin{equation}
|n|=wt,\quad w=\mathrm{const},\quad t\to\infty
,\quad t \in\mathbb R
,\quad n\in\mathbb R
\label{sol-front1}
\end{equation}
considering $n$ as a continuum spatial variable.
Here the meaning of the quantity 
\begin{equation}
0<w<1
\label{w-0-1}
\end{equation}
is the speed for the observation point.
This approach, which is known to us due to \cite{Slepyan1972} in context
of continuum problems, 
allows one to describe running waves, wave-fronts, and to
describe the wave-field as a whole, in comparison with the evaluation of the
corresponding asymptotics at a fixed position.
Note that the cases $n=0$ and $n=t$
are considered separately, in Sect.~\ref{bou-contrib} and
Sect.~\ref{sec-onfront}, respectively.
Denote 
\begin{gather}
\phi=w\arccos \frac{2-\Omega^2}2-\Omega
,
\\
\mathscr A^\mathrm{pass}(\Omega)\=\mathscr D^{-1}(\Omega)\=
\frac1{-(\DM)\Omega-\I\sqrt{4-\Omega^2}}.
\label{A-D-def}
\end{gather}
Thus,
\begin{gather}
\Idva_{wt}=-\frac{\I }{2\pi}\int_{0}^{2}
\mathscr A^\mathrm{pass}(\Omega)\EXP{\I \phi(\Omega) t}
\,\d \Omega.
\label{I-2-pass-band-1}
\end{gather}
The critical points for $I^\mathrm{pass}_{wt}$ are the stationary
point for $\phi$, where 
\begin{equation}
\phi'_\Omega=0, 
\label{stat-def}
\end{equation}
and the singular end-point $\Omega=\Omega_\ast=2$. Note that the
contribution $\Idva_{wt}(0)$ from the end-point $\Omega=0$ totally compensates by the
complexly conjugated integral over $(-2,0)$, see term $\cc$ in 
Eq.~\eqref{dot-u-fourier-gen}.

To estimate the contribution from the singular end-point $\Omega=2$, we
consider the behaviour of the amplitude 
$\mathscr A^\mathrm{pass}(\Omega)$
and the phase 
$\phi(\Omega)$
at $\Omega\to2-0$. 
One has
\begin{gather}
\begin{multlined}       
\mathscr A^\mathrm{pass}(\Omega)
=
\mathscr A^\mathrm{pass}_0
+
\mathscr A^\mathrm{pass}_{1/2}\sqrt{2-\Omega}+O(2-\Omega)
\\\qquad=-\frac1{2(\DM)}+\frac\I{2(\DM)^2}\sqrt{2-\Omega}+O(2-\Omega),
\end{multlined}
\label{A-pass-expa}
\\
\phi(\Omega)=\pi w-2-2w\sqrt{2-\Omega}+(2-\Omega)+o(2-\Omega).
\label{phi-pass-expa}
\end{gather}
{Taking into account the {Erd\'elyi} lemma
(see~Appendix~\ref{sec-erdelyi}), wherein $\alpha=1/2$, $\beta=1$, one gets that the
contribution $\Idva_{wt}(\Omega_\ast)$ from the end-point
$\Omega=\Omega_\ast\equiv2$ is $O(t^{-2})$ if 
\eqref{w-0-1} is fulfilled.}

\begin{remark}  
In the case $w=0$, which is considered in 
Sect.~\ref{bou-contrib}, the last estimation is $O(t^{-1})$ due to
Eq.~\eqref{phi-pass-expa}. This corresponds to the choice $\alpha=1$, $\beta=1$
in the {Erd\'elyi} lemma.
\end{remark}

The stationary point is the solution 
of Eq.~\eqref{stat-def}:
\begin{equation}
w\left(\arccos \frac{2-\Omega^2}2\right)'_\Omega-1=0.
\end{equation}
Thus, the expression for the stationary point $\Omega_\asst$ is
\begin{equation}
\Omega_\asst=2\sqrt{1-w^2}, 
\label{stat-point-pass-band-1}
\end{equation}
and
\begin{equation}
0<\Omega_\asst<2.
\label{stat-point-pass-band-1-ineq}
\end{equation}
One can see that the stationary point exists and unique for all $w$ in
interval \eqref{w-0-1}.
Put
\begin{gather}
\phi_\asst\=\phi(\Omega_\mathrm{s})=
w\arccos(2w^2-1)-2\sqrt{1-w^2}
=2(w\arccos w-\sqrt{1-w^2} ).
\label{phi_ast-expr}
\end{gather}
It can be shown that the stationary point $\Omega_{\mathrm s}$ is not a
degenerate one:
\begin{gather}
\phi''=\frac{\Omega w}{4(1-\frac{\Omega^2}{4})^{3/2}}>0,
\\
\phi''(\Omega_\mathrm{s})=
\frac{\sqrt{1-w^2}}{2w^2}
,
\label{phi-2deriv-abs}
\end{gather}
provided that inequalities
\eqref{w-0-1}
and 
\eqref{stat-point-pass-band-1-ineq}
are fulfilled. Now we can apply the formula for the principal term of 
the contribution 
from a non-degenerate stationary point $\Omega_\mathrm{s}$.\footnote{%
This formula follows from the Erd\'elyi lemma for $\alpha=1$, $\beta=2$
(see Appendix~\ref{sec-erdelyi}), see, e.g., \cite{Fedoryuk1977,temme2014}.} 
One obtains 
%
the following asymptotics:
\begin{multline}
\VF_{wt}^\mathrm{pass}=
\Idva_{wt}(\Omega_\mathrm{s})+
\Idva_{wt}(\Omega_\ast)+\cc+O(t^{-\infty})
=
\Idva_{wt}(\Omega_\mathrm{s})
+\cc+O(t^{-2})
\\=-\frac{\I }{\sqrt{2\pi|\phi''(\Omega_\asst) |t}}\,
\frac{\EXP{\I(\phi_\asst t+\frac{\pi}{4})}}
{\mathscr D(\Omega_\mathrm{s})}+\cc+O(t^{-3/2}).
\label{cont-pass}
\end{multline}
Using Eq.~\eqref{A-D-def} one gets
\begin{gather}  
\Re \A
=-(\DM)\Omega_\asst=
-2(\DM)\sqrt{1-w^2}
,\\
\Im \A
=-\sqrt{4-{\Omega_\asst^2}}
=-2w
,\\
\begin{multlined}       
|\A|^2=
4\big((\DM)^2(1-w^2)+w^2\big).
\end{multlined}
\end{gather}
Hence, one can obtain:
\begin{multline}
\VF_{wt}^\mathrm{pass}=
-\frac{2}{\sqrt{{2\pi}\phi''_\asst t}}
\frac{1}{|\A|^2}
\bigg(\Im \A\cos\Big(\phi_\asst t+\frac{\pi}{4}\Big)
- \Re \A
\sin\Big(\phi_\asst t+\frac{\pi}{4}\Big)
\bigg)
+O(t^{-3/2})
\\
=-\frac{1}{\sqrt{\pi t}}
\frac{
w\Big(-w\cos\Big(\phi_\asst
t+\frac{\pi}{4}\Big)
+
(\DM)\sqrt{1-w^2}
\sin\Big(\phi_\asst t+\frac{\pi}{4}\Big)
\Big)
}{{(1-w^2)^{1/4}}\big((\DM)^2(1-w^2)+w^2\big)}
+O(t^{-3/2})
\\=
\frac{1}{\sqrt{\pi t}}
\frac{ w\cos\left(\phi_\asst t + \tphi + \frac{\pi}{4}\right) }
{{(1-w^2)^{1/4}}\big((\DM)^2(1-w^2)+w^2\big)^{1/2}}
+O(t^{-3/2})
,
\label{I1+I2-w}
\end{multline}
where 
\begin{equation}
\tphi=\arctan\frac {\Re \mathscr D}{\Im \mathscr D}
=
\arctan 
\frac
{(\DM)\sqrt{1-w^2}}
{w}.
\label{psi-def}
\end{equation}

Now we return to variables $n$, $t$, and substitute $w=|n|/t$ into the last
expression. This yields:
\begin{multline}
\VF_{|n|=wt}^\mathrm{pass}=-\frac{
|n| }{\sqrt{\pi}(t^2-n^2)^{1/4}
\big((\DM)^2(t^2-n^2)+n^2\big)
}
\\
\times
\Big(
-|n|\cos\Big(\phi_\asst t+\frac{\pi}{4}\Big)
+
(\DM)\sqrt{t^2-n^2}\sin\Big(\phi_\asst t+\frac{\pi}{4}\Big)
\Big)
+O(t^{-3/2})
\\=
\frac{|n|}
{\pi^{1/2}(t^2-n^2)^{1/4}
\left( (\DM)^2(t^2-n^2)+n^2\right)^{1/2}
}
\cos\left(\phi_\asst t + \tphi + \frac{\pi}{4}\right)
+O(t^{-3/2})
,
\label{I_1+I_2}
\end{multline}
where according to Eqs.~\eqref{phi_ast-expr}, \eqref{psi-def}
\begin{gather}
\phi_\asst=
2\left(
\frac {|n|}t \arccos \frac {|n|}t
-\frac{\sqrt{t^2-n^2}}{t}
\right),
\label{phis-def-n}
\\
\tphi
=
\arctan 
\frac
{(\DM)\sqrt{t^2-n^2}}
{|n|}
%
.
\label{psi-def-n}
\end{gather}


\subsection{The contribution from the stop-band}
\label{sband-co}

One has:
\begin{equation}
I_n^\mathrm{stop}=
-\frac{\I }{2\pi}\int_{2}^{\infty}\pFO
\frac{(-1)^{|n|}\,\Omega\,\EXP{-b(\Omega)|n|-\I\Omega t}}
{-m\Omega^2+2\EXP{-b(\Omega)}+2}
\, \d \Omega.
\label{I-2-stop-band}
\end{equation}
The contribution from the stop-band describes non-propagating exponentially
vanishing as $n\to\infty$ waves.
Therefore, we estimate the large-time asymptotics 
of the right-hand side of 
Eq.~\eqref{I-2-stop-band}
at a fixed spatial position $n$.

While evaluating the contribution from the stop-band, we have two
qualitatively different cases, namely, the case of a heavy defect ($m>1$)
and the case of a light defect ($m<1$). Indeed, the localized mode
exists in the system if and only if $m<1$ (see Sect.~\ref{sec-spectral-p}). 
The frequency of localized
oscillation is a root of frequency equation 
\eqref{freq-eq-complex-wn-al0-0}.
Thus, the pole $\Omega=\Omega_0$ exists inside the integration
interval $\mathbb S_+$ if and only if $m<1$. 

In the case $m>1$ there is no pole and integral \eqref{I-2-stop-band} exists
in the classical sense. The only critical point for the integral is the cut-off
frequency $\Omega_\ast=2$:
\begin{equation}
I_n^\mathrm{stop}=I_n^\mathrm{stop}(\Omega_\ast)+O(t^{-\infty}).
\end{equation}
{A rough estimation $I_n^\mathrm{stop}(\Omega_\ast)=O(t^{-1})$ can be obtained from
the Erd\'elyi lemma with $\alpha=1$, $\beta=1$}%
\footnote{In Sect.~\ref{bou-contrib} we discuss the contribution
$I_0^\mathrm{stop}(\Omega_\ast)$ in more details.}.
Thus, in the case of a heavy defect $m>1$ one has
\begin{equation}
\VF_n^\mathrm{stop}=O(t^{-1}).
\label{stop-empty}
\end{equation}

In the case $m<1$ there is the pole $\Omega_\mathrm{s}\in\mathbb S_+$ and
integral \eqref{I-2-stop-band} does not exist
in the classical sense. In the latter case the integral 
in the right-hand side of \eqref{I-2-stop-band} should be considered as
the Fourier transform for a generalized function. To get
the possibility to deal with the right-hand side of \eqref{I-2-stop-band} as
an ordinary Lebesgue integral we use the limit absorption principle.
We add the dissipative viscous term into governing equation 
\eqref{chain-eq-basic-al0},
repeat all the
calculations and find the root of the denominator for the right-hand side of 
Eq.~\eqref{chain-eq-basic-al0}
modified in such a way (see \cite{Gavrilov1999jsv} for details). Then we consider a limiting case of zero dissipation
to define the positions of the poles with respect to the real axis. One can
show that the poles are shifted into the lower half-plane of the complex
plane: $\Omega=\pm\Omega_0-\I0$. To calculate the contribution
$I_n^\mathrm{stop}(\Omega_0)$
we apply Sokhotski–Plemelj theorem  for the real line \cite{Vladimirov1971}:
\begin{equation}
\int_{-\infty}^{\infty}\frac{\bar f(\Omega)
}{\Omega\pm \I0}\,\d\Omega=
\mp \I\pi f (0)+
\PV\int_{-\infty}^{\infty}\frac{\bar f(\Omega)
}{\Omega}\,\d\Omega,
\label{S-P}
\end{equation}
and the following asymptotic formula \cite{Fedoryuk1977} for the asymptotics of the Cauchy principal value
for the integral
\begin{gather}
\PV\int_{-\infty}^{\infty}\frac{f(\Omega)\,e^{-\I\Omega t}}{\Omega}\,\d\Omega=
-\pi \I\,f(0)
+O(t^{-\infty}).
\label{PV}
\end{gather}
Here {$\bar f(\Omega), f(\Omega)\in C_0^\infty$}.
Now one can put 
\begin{equation}
\bar f(\Omega)=f(\Omega)\, \EXP{-\I\Omega t}
\end{equation}
and use Eq.~\eqref{S-P}. This 
yields
\begin{gather}
\int_{-\infty}^{\infty}\frac{f(\Omega)\,e^{-\I\Omega t}}{\Omega-\I0}\,d\Omega=
O(t^{-\infty}),
\label{0-def}
\\
\int_{-\infty}^{\infty}\frac{f(\Omega)\,e^{-\I\Omega t}}{\Omega+\I0}\,d\Omega=
-2\pi \I\,f(0)+O(t^{-\infty}).
\label{non-an}
\end{gather}
Using Eq.~\eqref{non-an}, one gets
\begin{multline}
\int_{-\infty}^{\infty}\frac{f(\Omega)\,e^{-\I\Omega t}}{\Omega-\Omega_0+\I0}\,d\Omega=
-2\pi \I\,f(\Omega_0)\,e^{-\I\Omega_0 t}+O(t^{-\infty})
\\
=-2\pi \I\,\Res\big(F(\Omega),\Omega_0\big)\,e^{-\I\Omega_0 t}+O(t^{-\infty}).
\label{Res-def}
\end{multline}
The last equality is correct provided that 
$F(\Omega)=f(\Omega)/(\Omega-\Omega_0)$ is an analytic function in a neighbourhood of
$\Omega_0$.

Applying Eq.~\eqref{Res-def} one obtains
\begin{gather}
I_n^\mathrm{stop}(\Omega_0)=
I_0^\mathrm{stop}(\Omega_0)
(-1)^{|n|}\EXP{-b(\Omega_0)|n|},\\
I_0^\mathrm{stop}(\Omega_0)=- {\Omega_0} 
\Res\left(\frac{1}{-m\Omega^2+2\EXP{-b}+2},{\Omega_0}
\right)\EXP{-\I{\Omega_0}t}+O(t^{-\infty}),
\label{cont-Omega0}
\end{gather}
where
\begin{multline}
\Res\left(\frac{1}{-m\Omega^2+2\EXP{-b}+2},\Omega_0
\right) 
=
\frac{1}{-2 m\Omega-2b'_{\Omega}\EXP{-b}}
\\
=\frac{\sinh b}{\Omega_0(-m\EXP b+(m-2)\EXP{-b})}
=-\frac{(\DM)\Omega_0}{2\big(m(m-2)\Omega_0^2+2\big)}.
\label{Res-calc}
\end{multline}
Here, to simplify the left-hand side of \eqref{Res-calc} we have used relation
\begin{equation}
b'_{\Omega}=\frac{\Omega}{\sinh b(\Omega)},
\end{equation}
which follows from Eq.~\eqref{eq-complex-wn-1-al0}.
Also, we have used Eq.~\eqref{freq-eq-complex-wn-al0} to calculate $\exp(-b)$ and 
Eq.~\eqref{eq-complex-wn-2-al0} to calculate $\sinh b$.
Using Eqs.~\eqref{freq-eq-complex-wn-al0}, \eqref{loc-freq-m<1-zero}
one finally obtains:
\begin{equation}
\VF_n^\mathrm{stop}=\frac{(\DM)(-1)^{|n|+1}m^{|n|-1}}{(2-m)^{|n|+1}}
\EXP{-\I\Omega_0 t}
+I_n^\mathrm{stop}(\Omega_\ast)+O(t^{-\infty})+\cc,
\end{equation}
where $I_n^\mathrm{stop}(\Omega_\ast)=O(t^{-1})$.
Finally,
\begin{equation}
\VF_n^\mathrm{stop}=\frac{2(\DM)(-1)^{|n|+1}m^{|n|-1}}{(2-m)^{|n|+1}}
\cos\Omega_0 t
+O(t^{-1}).
\label{dotu_k-st-band-1}
\end{equation}
{Formula \eqref{dotu_k-st-band-1} describes non-vanishing localized
oscillation,} which exists in the case of a light defect $m<1$. For the first
time this result was obtained in \cite{Kashiwamura1962}.

\subsection{The case $n=0$: the contribution from the cut-off frequency} 
\label{bou-contrib}
Consider the case $w=0$, which corresponds
to a fixed position $n=0$. In this case the stationary point
$\Omega_\mathrm{s}$ collocates with  the cut-off frequency:
($\Omega_\mathrm s\to\Omega_\ast-0$ as $w\to+0$), and disappears.
Accordingly, the term of order $t^{-1/2}$ of the asymptotics for 
$\VF_{0}^\mathrm{pass}$ vanishes, see 
Eq.~\eqref{I1+I2-w}. 
In this section we estimate the principal term for
\begin{equation}
I_0(\Omega_\ast)+\cc=
I_0^\mathrm{pass}(\Omega_\ast)
+
I_0^\mathrm{stop}(\Omega_\ast)
+\cc
\end{equation}

Due to Eqs.~\eqref{contrib}, \eqref{I-2-pass-band-1} one has
\begin{equation}
\Idva_{0}(\Omega_\ast)=
-\frac{\I }{2\pi}\int_{-\infty}^{2}
\chi_2(\Omega)\mathscr A^\mathrm{pass}(\Omega)\EXP{-\I\Omega t}
\,\d \Omega,
\label{I0-pass}
\end{equation}
where 
$\chi_2(\Omega)$ is a neutralizer (see
Remark~\ref{neu-def}),
and expansion 
\eqref{A-pass-expa}
is valid for $\Omega\to\Omega_\ast-0$. On the other hand, 
due to Eqs.~\eqref{contrib}, \eqref{I-2-stop-band}
\begin{gather}
I_0^\mathrm{stop}(\Omega_\ast)
=-\frac{\I }{2\pi}\int_{2}^{\infty}
\chi_2(\Omega)
\mathscr A^\mathrm{stop}(\Omega)\EXP{-\I\Omega t}
\,\d \Omega
,
\label{I0-stop}
\\
\mathscr A^\mathrm{stop}\=\frac{\Omega}{-m\Omega^2+2\EXP{-b}+2}.
\end{gather}
One can show that
\begin{multline}
\mathscr A^\mathrm{stop}=
\mathscr A^\mathrm{stop}_0
+
\mathscr A^\mathrm{stop}_{1/2}\sqrt{\Omega-2}+o(\sqrt{\Omega-2})
\\=-\frac{1}{2(\DM)}+\frac{1}{2(\DM)^2}
\sqrt{\Omega-2}+o(\sqrt{\Omega-2}),\quad \Omega\to2+0.
\end{multline}
Since $\mathscr A_0^\mathrm{pass}=\mathscr A_0^\mathrm{stop}$,
\begin{equation}
\underbrace{\int_{-\infty}^{2}
\chi_2(\Omega)\mathscr A_0^\mathrm{pass}(\Omega)\,\EXP{-\I\Omega t}
\,\d \Omega}_{J}
+
\int_{2}^{\infty}
\chi_2(\Omega)
\mathscr A_0^\mathrm{stop}(\Omega)\,\EXP{-\I\Omega t}
\,\d \Omega=O(t^{-\infty}).
\end{equation}
Applying the Erd\'elyi lemma with $\alpha=1,\ \beta=3/2$ yields
\begin{multline}  
I_0^\mathrm{pass}(\Omega_\ast)=
J+
\frac1{2\pi}\int_0^{\infty} 
\big(|\mathscr A^\mathrm{pass}_{1/2}|\sqrt\mu
+o(\sqrt\mu)
\big)
\EXP{\I(\mu-2)t}\,\d\mu+\cc+O(t^{-\infty})
\\=
2\Re J+2\Re
\frac{|\mathscr A^\mathrm{pass}_{1/2}|
\Gamma\left(\frac32\right)\EXP{\I\left(2t-\frac{3\pi}4\right)}}
{2\pi\, t^{3/2}}
+ o(t^{-3/2})
\\=
2\Re J+\frac{\cos\left(2t-\frac{3\pi}4\right)}{4\sqrt\pi(\DM)^2\, t^{3/2}}
+ o(t^{-3/2}),
\end{multline}
\begin{multline}  
I_0^\mathrm{stop}(\Omega_\ast)=
-J
-\frac\I{2\pi}\int_0^\infty \big(\mathscr A^\mathrm{stop}_{1/2}\sqrt\mu
+o(\sqrt\mu)\big)
\EXP{-\I(2+\mu)t}\,\d\mu+\cc+O(t^{-\infty})
\\=-2\Re J+
2\Re
\frac
{\mathscr A^\mathrm{stop}_{1/2}
\Gamma\left(\frac32\right)\EXP{\I\left(2t+\frac{3\pi}4+\frac\pi2\right)}}
{2\pi\, t^{3/2}}
+ o(t^{-3/2})
\\=
-2\Re J+
\frac{\cos\left(2t-\frac{3\pi}4\right)}{4\sqrt\pi(\DM)^2\, t^{3/2}}
+ o(t^{-3/2})
.
\end{multline}
Here relation 
\begin{equation}
\Gamma\left(\frac32\right)=\frac{\sqrt\pi}2
\end{equation}
is taken into account.
Finally, 
{one has}
\begin{equation}
\VF_0=I_0^\mathrm{stop}(\Omega_0)H(1-m)+
\frac{\cos\left(2t-\frac{3\pi}4\right)}{2\sqrt\pi(\DM)^2\, t^{3/2}}
+ o(t^{-3/2})
.
\label{rubin-f}
\end{equation}
Here $I_0^\mathrm{stop}(\Omega_0)$ is defined by Eqs.~\eqref{cont-Omega0},
\eqref{Res-calc} in the case of a light defect.
Formula 
\eqref{rubin-f} was previously obtained in \cite{Rubin_1963}.


\subsection{The case $|n|=t$}
\label{sec-onfront}

One has $\Omega_\mathrm{s}\to+0$ as $w\to1$, therefore for $w=1$ two
stationary points collocate:
\begin{equation}
\phi''\big|_{\Omega_\mathrm{s}=0}=0,\qquad
\phi'''\big|_{\Omega_\mathrm{s}=0}=\frac14.
\end{equation}
The principal term of the asymptotics on the front $w=1$ ($n=\pm t$) 
can now be calculated by the Erd\'elyi
lemma, wherein $\alpha=1/3$, $\beta=1$:
\begin{multline}
\VF^\mathrm{pass}_{\pm t}=-\frac{\I }{2\pi}\int_{-2}^{2}\chi_0(\Omega)
\big(\mathscr A^\mathrm{pass}(0)+o(1)\big)
\EXP{\I t\left(\frac{ \phi'''(0)\Omega^3}6+o(1)\right)}
\,\d \Omega+O(t^{-\infty})\\=
\frac{\Gamma(\frac13)}{2\cdot 3^{1/6}\pi}\,t^{-1/3}+o\big(t^{-1/3}\big).
\label{v-onfront}
\end{multline}
Thus, the principal term for the particle velocity on the leading front $|n|=t$ 
does not depend on $m$, and  coincides with the
corresponding result for the uniform chain. 

More detailed uniform asymptotics
describing the behaviour of 
$\VF_{n\simeq |t|}^\mathrm{pass}$ in a certain neighbourhood of the leading wave
front $n=\pm t$
can be obtained by considering collocations
of the stationary points $\pm \Omega_s\simeq0$. {This solution can be expressed
in terms of the Airy function \cite{Fedoryuk1977,temme2014}.}

\subsection{The case $|n|>t$}
\label{n>t}
In the case \eqref{sol-front1} wherein $w>1$ 
the only critical points for $I^\mathrm{pass}_{wt}$ is
the singular end-point $\Omega=\Omega_\ast=2$. The corresponding contribution
can be estimated in the same way as in Sect.~\ref{sect-pass} {as $O(t^{-2})$}. 
On the other hand, the integral $I^\mathrm{stop}_{wt}$ is not a Fourier
integral any more, and thus it cannot be evaluated by the stationary phase
method. We expect that 
\begin{equation}
\VF_{wt}=
I^\mathrm{pass}_{wt}+
I^\mathrm{stop}_{wt}+\cc=
O(t^{-\infty}) 
\label{k>t}
\end{equation}
as well as it is observed in the uniform chain, where this result follows from
the exact analytic solution in terms of Bessel functions \cite{Olver1997,Sokolov2021}. Note that
numerical calculations are in agreement with this hypothesis.
Nevertheless, there we do not see an easy way to prove 
asymptotics \eqref{k>t} in the case $m\neq1$ basing on integral representation
\eqref{dot-u-fourier-gen}.

\subsection{Summary: the approximate solution}
\label{sect-before-front}
Let us summarize the results obtained in Sect.~\ref{sec-non}.
According to
Eqs.~\eqref{I_1+I_2}, \eqref{dotu_k-st-band-1}
{the approximate solution $\VA_n$: $\VA_n\simeq \VF_n$}, which we use in
Sect.~\ref{sec-num}
to investigate thermal motions, is
\begin{align}
&\VA_n\=\VA_n^\mathrm{pass},&&m>1;
\label{v-pass}
\\
&\VA_n\=\VA_n^\mathrm{pass}+\VA_n^\mathrm{stop},&& 0< m<1;
\label{v-pass-stop}
\end{align}
where
\begin{gather}
\VA_n^\mathrm{pass}\=
\frac{H(t-|n|)\, |n|}
{\pi^{1/2}(t^2-n^2)^{1/4}
\left( (\DM)^2(t^2-n^2)+n^2\right)^{1/2}
}
\cos\left(\phi_\asst t + \tphi + \frac{\pi}{4}\right),
\label{v-decay}
\\
{v}_n^\mathrm{stop}=\frac{2(\DM)(-1)^{|n|+1}m^{|n|-1}}{(2-m)^{|n|+1}}
\cos\Omega_0 t.
\label{v-loc}
\end{gather}
Here $\phi_\mathrm{s}$ and $\psi$ are defined by Eqs.~\eqref{phis-def-n} and
\eqref{psi-def-n}, respectively. 

\begin{remark}  
\label{remark-vn}
By construction (see Eq.~\eqref{sol-front1}),
quantity $\VA_n^\mathrm{pass}$ in Eq.~\eqref{v-decay}
can be considered as a continuum one ($n\in \mathbb R$),
which is equal to the particle velocity of a particle with number $n$ for 
$n\in \mathbb Z$. On the other hand, $n\in \mathbb Z$ in Eq.~\eqref{v-loc} for 
$\VA_n^\mathrm{stop}$,
since for the localized mode neighbouring particles oscillate in counter
phase.
\end{remark}

Formula \eqref{v-pass} 
has an exact asymptotic meaning. For $|n|<t$, $n\neq0$, the right-hand
side coincides with the principal term of order $t^{-1/2}$ for the asymptotics
on the moving front $|n|=wt$. For $|n|>t$ we expect that the 
corresponding asymptotics is $O(t^{-\infty})$, see Sect.~\ref{n>t}.
For $w=0$ (or $n=0$) the principal term is of order $t^{-3/2}$ and defined by
Eq.~\eqref{rubin-f}. For $w=1$ (or $|n|=t$) the principal term is of order $t^{-1/3}$ and defined by
Eq.~\eqref{v-onfront}.

Formula 
\eqref{v-pass-stop}
is not asymptotically correct, since two terms in the right-hand side are
asymptotics calculated at a moving position and at a fixed one, respectively.
Formula \eqref{v-decay} describes decaying with the time propagating component of the
wave-field,
which exists for any positive value of~$m$. 
{Formula \eqref{v-loc} describes non-vanishing localized oscillation, which
exists in the case of a light defect $0<m<1$.}

\section{Numerics}
\label{sec-num}
The obtained analytical results were compared with the results of the 
of numerical integration of $N=2n+1$
ODE 
\eqref{chain-eq-basic-al0} with {periodic conditions at the ends of the chain}.\footnote{The specific form of this boundary conditions is not very important
in our calculations, since we take large enough $N$.}

\begin{figure}[htb]
\centering\includegraphics[width=0.9\textwidth]{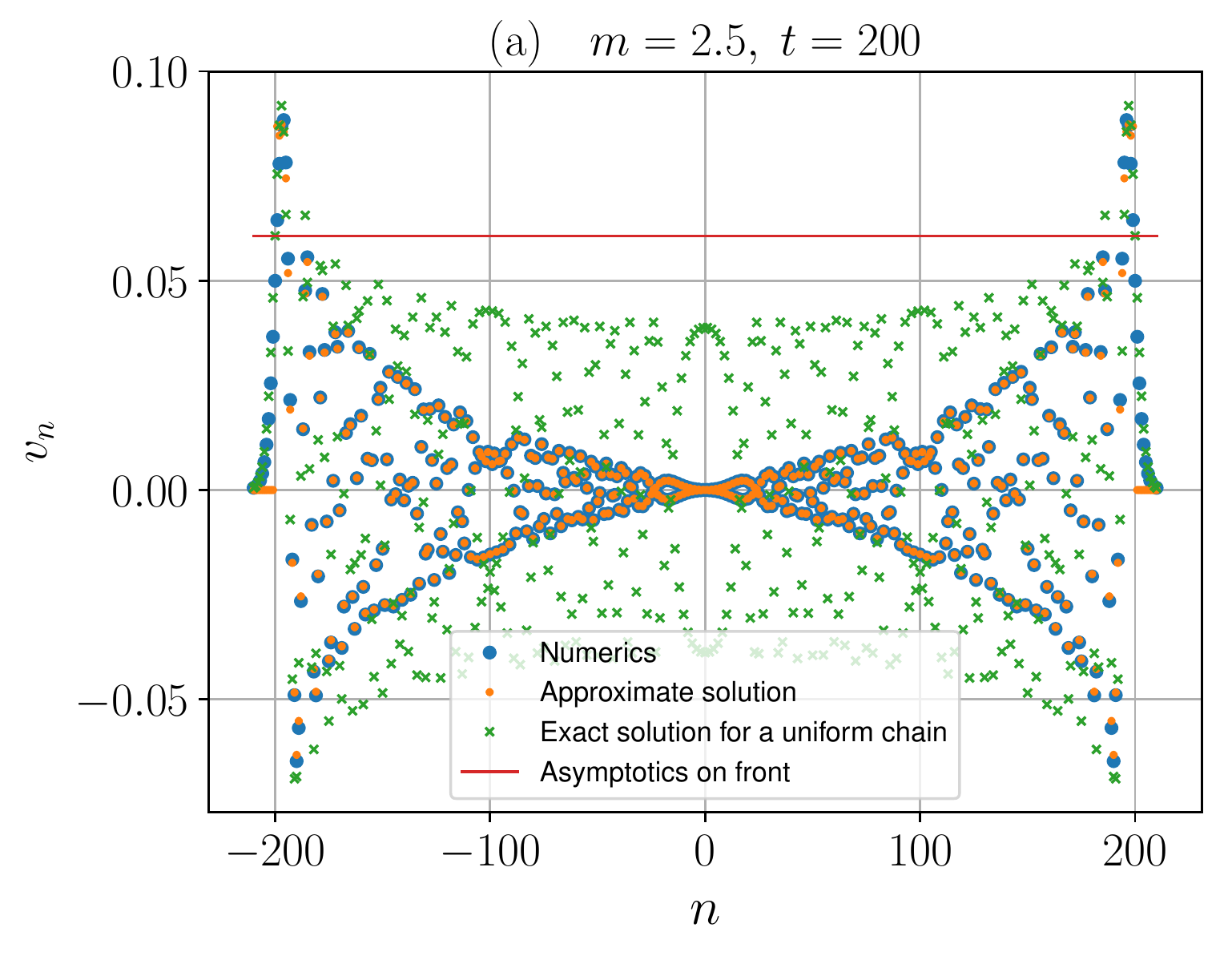}
\centering\includegraphics[width=0.9\textwidth]{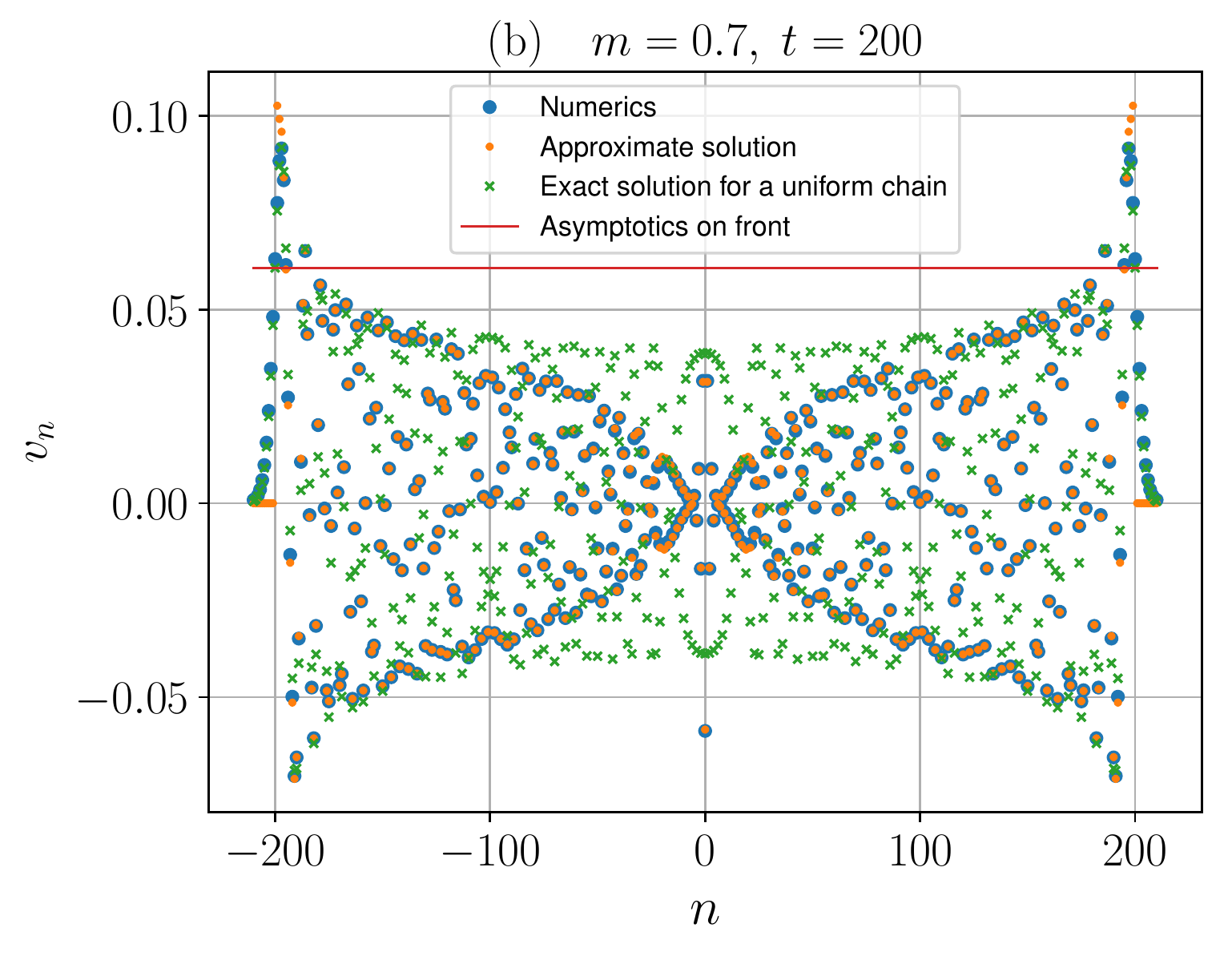}
\caption{Comparing the approximate solution $v_n$ in the form of 
Eqs.~\eqref{v-pass}--\eqref{v-loc} and numerical solution (particle velocities
versus particle numbers). (a) The case of
heavy defect, (b) the case of a light defect. The green crosses correspond
to the corresponding exact solution \eqref{Sro-bessel} for a uniform chain.
The red solid line corresponds to the value of $v_n$ given by 
Eq.~\eqref{v-onfront} on the leading front $|n|=t$}
\label{Vk.pdf}
\end{figure}

In Fig.~\ref{Vk.pdf} we compare the approximate solution $v_n$ in the form of 
Eqs.~\eqref{v-pass}--\eqref{v-loc} and the corresponding numerical solution in
the cases of a heavy defect and a light one (particle velocities
versus particle numbers). 
We also demonstrate the corresponding exact solution \eqref{Sro-bessel} for a
uniform chain at the same plot. One can clearly observe the effects of anti-localization 
(for the both heavy and light defects) and
localization (for the case of a light defect). The obtained analytical
solution is in a good agreement with the numerical one everywhere excepting
the wave front $|n|=t$. Remember, that solution    
\eqref{v-pass}--\eqref{v-loc} is not valid by construction near $|n|=t$ (see
Sect.~\ref{sec-onfront}). At the leading wave front asymptotics \eqref{v-onfront}
should be used instead. The corresponding value is also marked at the plot to
demonstrate a good agreement.

\begin{figure}[htb]
\centering\includegraphics[width=0.9\textwidth]{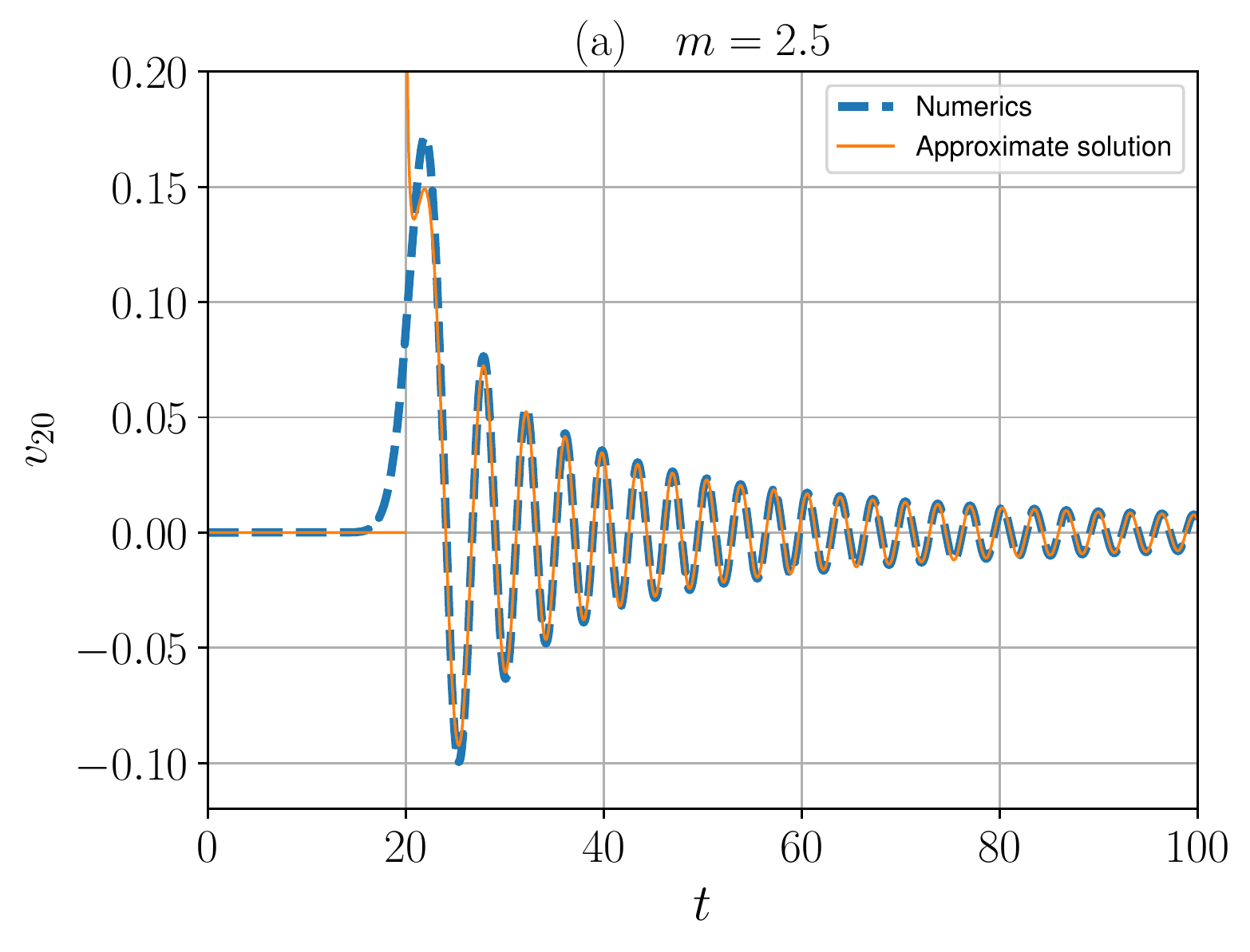}
\centering\includegraphics[width=0.9\textwidth]{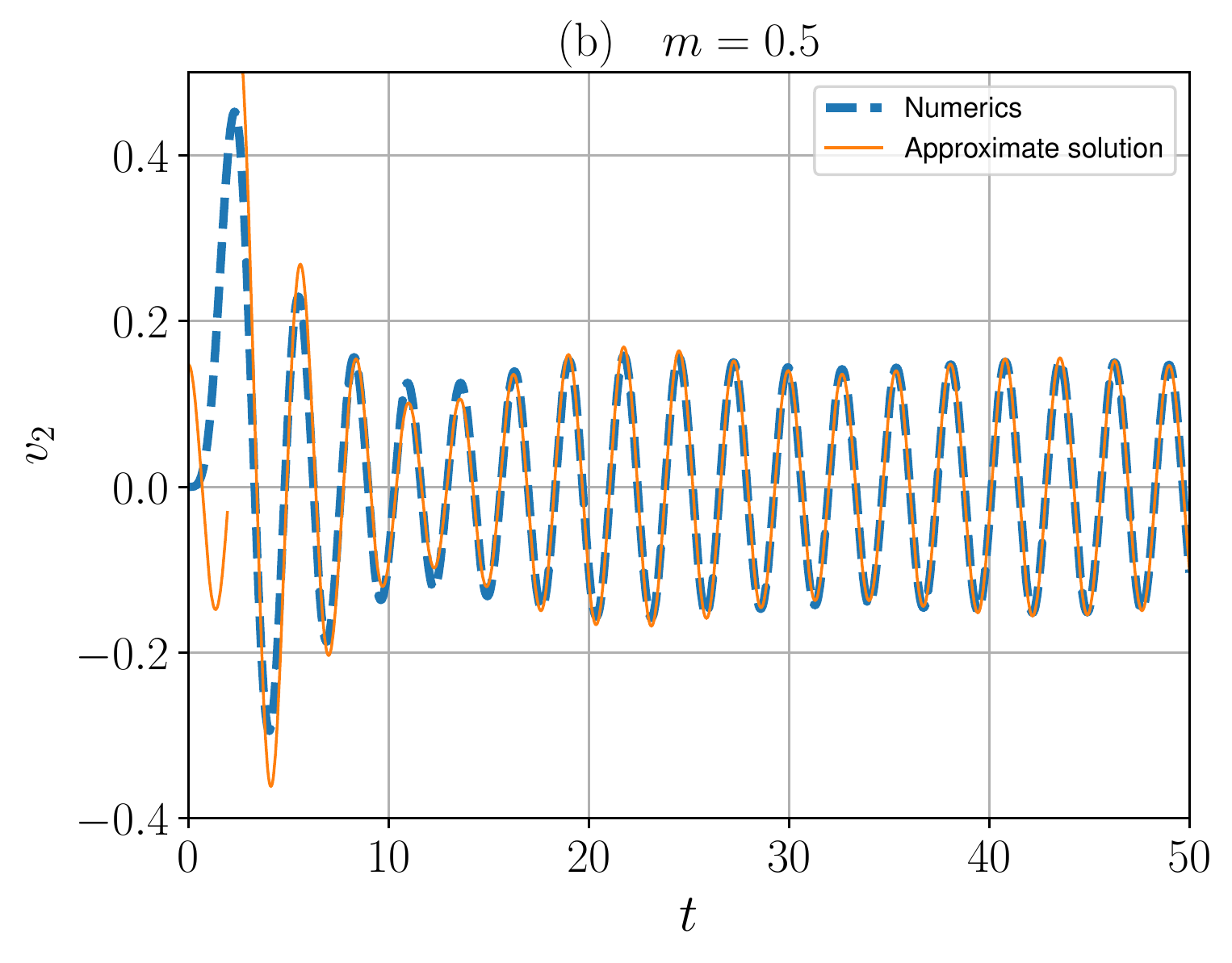}
\caption{Comparing the approximate solution $v_n$ in the form of 
Eqs.~\eqref{v-pass}--\eqref{v-loc} and numerical solution (particle velocity
versus time). (a) The case of
a heavy defect, (b) the case of a light defect}
\label{Vt.pdf}
\end{figure}

\begin{figure}[htb]
\centering\includegraphics[width=0.9\textwidth]{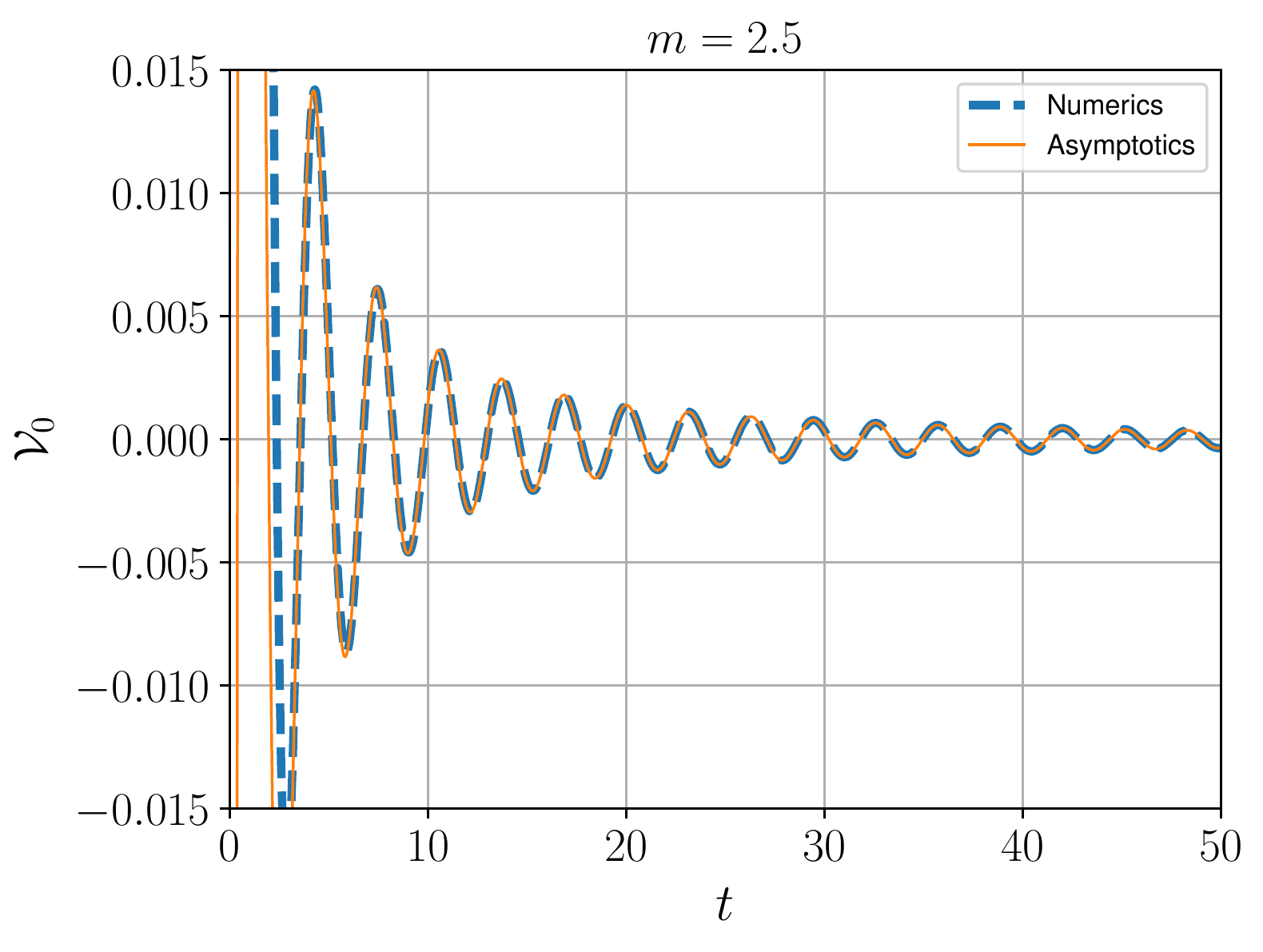}
\caption{
Comparing the asymptotics for $\mathcal V_0$ in the form of 
Eq.~\eqref{rubin-f} and numerical solution 
(particle velocity
versus time) in the case of heavy defect}
\label{cut-off.pdf}
\end{figure}

In Fig.~\ref{Vt.pdf} we provide the analogous comparison for particle velocities
versus time. One can observe the vanishing oscillation in the case of a heavy
defect and undamped non-vanishing oscillation in case of a light defect.
Finally, in Fig.~\ref{cut-off.pdf} we compare in the
case of a heavy defect the asymptotic solution
\eqref{rubin-f} for $\VF_0(t)$ with the corresponding numerical results and
demonstrate a good agreement.

\afterpage{\clearpage}
\section{Discussion: the cases of a light defect, a heavy one, and a
uniform chain}
\label{sect-disc}
Put $m=1$ in the approximate solutions for $m>1$ or $m<1$ (Eq.~\eqref{v-pass}
or Eq.~\eqref{v-pass-stop}, respectively). 
This yields
\begin{equation}
\VA_n=\VA_n^\mathrm{pass}=
\frac{H(t-|n|)\, |n|}
{\pi^{1/2}(t^2-n^2)^{1/4}
|n|
}
\cos\left(\phi_\asst t +  \frac{\pi}{4}\right).
%
\label{asymp-simple-chain1}
\end{equation}
For $n\neq0$ we can drop $|n|$ in the numerator and the denominator and get
the expression 
\begin{equation}
\VA_n=
%
\frac{
{H(t-|n|)}
}{
\pi^{1/2}(t^2-n^2)^{1/4}
}
\cos\left(\phi_\asst t +  \frac{\pi}{4}\right).
\label{asymp-simple-chain}
\end{equation}
The last expression was previously obtained \cite{Gavrilov2022ijhmt} 
as the principal term for the asymptotics of the exact analytic
solution for the problem concerning a uniform chain. Thus, for
$n\not=0$ the approximate solution for $m>1$ or $m<1$
(Eq.~\eqref{v-pass} or
Eq.~\eqref{v-pass-stop}, respectively) transforms into the approximate solution
for the uniform chain.



{In the case of a heavy defect $m>1$ the approximate solution
\eqref{v-decay} is zero at $n=0$ (and this is not true for the uniform chain,
see 
Eq.~\eqref{asymp-simple-chain}).
Speaking more accurately, this fact means that the
principal term of the 
asymptotic solution for $w=0$ is of order $O(t^{-3/2})$ and described by
Eq.~\eqref{rubin-f}, whereas it is of order $O(t^{-1/2})$ in a uniform chain.
Such a result and the corresponding asymptotics \eqref{rubin-f}
was obtained in the paper by Rubin \cite{Rubin_1963} and is in agreement with results
obtained in \cite{Kashiwamura1962,Mueller1962}. Moreover, the similar result was obtained by
Kaplunov \cite{kaplunov1986torsional}, who considered a continuum system with a point inertial inclusion,
where a localized (trapped) mode
exists. We suggest to call this phenomenon ``anti-localization''. The physical
meaning of anti-localization is a tendency for the unsteady propagating
wave-field to avoid a neighbourhood of an inclusion or a defect.}

In the case of a light defect $0<m<1$ the non-stationary localized oscillation exists
and dominates over all other motions for small $n$. 
A part of the total energy of the chain is trapped near the defect. Note that
the propagating part of the wave-field is also anti-localized.

{In the case $m\simeq 1,\ m\neq1$ the obtained approximate solution 
\eqref{v-pass}--\eqref{v-loc}, as well as formula \eqref{rubin-f},
become applicable at small $n$ after a very large time.
This is because two or three critical
points, namely the stationary point $\Omega_\mathrm{s}$ defined by 
Eq.~\eqref{stat-point-pass-band-1}, the cut-off frequency
$\Omega_\ast=2$, and 
the pole $\Omega_0$ in the case $m<1$}:
\begin{equation}
\Omega_0=2+(m-1)^2+O\big((m-1)^3\big),\qquad m\to1-0,
\end{equation}
collocate as $n=0\ (w=0)$ and $m\to1-0$. To describe this case better one
needs
to consider the collocations and to obtain the corresponding uniform asymptotics. This
is a separate difficult problem, which is beyond the scope of the current
paper.

\begin{remark}  
\label{remark-rubin}
Put $|\DM|=1$ in Eq.~\eqref{v-decay}. This yields
\begin{equation}
\VA_n^\mathrm{pass}=
\frac{H(t-|n|)\, |n|}
{\pi^{1/2}(t^2-n^2)^{1/4}
t
}
\cos\left(\phi_\asst t +  \frac{\pi}{4}\right).
\label{asymp-simple-chain2}
\end{equation}
For $\frac{|n|}t\sim1$ the right-hand side of Eq.~\eqref{asymp-simple-chain2}
coincides with the right-hand side of Eq.~\eqref{asymp-simple-chain} obtained
for the uniform chain. The corresponding asymptotics
in the case $m=2$ was obtained by Rubin in \cite{Rubin_1963},
and, strictly speaking, it is incorrect, because \eqref{v-decay} is inapplicable
for $\frac{|n|}t\sim1$ 
due to singularities collocation
(see Sect.~\ref{sec-onfront}). 
This
result and formula \eqref{rubin-f} are only results\footnote{See formulae
(A36)--(A39) in \cite{Rubin_1963}.}
concerning to propagating
part of the wave-field, which  
obtained in the paper by Rubin
\cite{Rubin_1963}, who estimated the integral representation of the solution at
a fixed position. 
\end{remark}
%
%
%

\section{The thermal motion}
\label{sec-thermal}

In this section we investigate the properties of the thermal fundamental
solution $\TT_n$
defined by \eqref{thermal-fs}. To do this we use the approximation $v_n$ defined by 
Eqs.~\eqref{v-pass-stop}--\eqref{v-loc} for the fundamental solution $\VF_n$
and obtain:
\begin{equation}
\TT_n\simeq
2m m_n \VA_n^2(t).
\label{TT-app}
\end{equation}

At first, consider the case $m>1$. One has $\VA_n=\VA_n^\mathrm{pass}$ due to 
Eq.~\eqref{v-pass}, where $\VA_n^\mathrm{pass}$ is defined by 
Eq.~\eqref{v-decay} for $n\in \mathbb R$ (see Remark~\ref{remark-vn}).
Since $\VA_0^\mathrm{pass}=0$, 
the right-hand side of Eq.~\eqref{TT-app} can be in a natural way rewritten 
as a smooth continuum quantity
\begin{equation}
\TT_n\simeq
2m  \VA_n^2(t),\quad n\in\mathbb R.
\label{TT-app-mod}
\end{equation}
Thus,
\begin{equation}        
\TT_n
\simeq
\TT_n^\mathrm{pass}
\=
\frac{2m  |n|^2
\cos^2
\left( \phi_{\mathrm{s}} t+\tphi+\frac{\pi}{4} \right)
\Hf
}
{\pi(t^2-n^2)^{1/2}
\left( (\DM)^2(t^2-n^2)+n^2 \right)}
=
\bar\TT_n^\mathrm{pass}
+
\hat\TT_n^\mathrm{pass},
\qquad n\in \mathbb R,
\label{T-pass-def}
\end{equation}
where
\begin{gather}        
\bar\TT_n^\mathrm{pass}
\=
\frac{mn^2 \Hf}
{\pi(t^2-n^2)^{1/2}\big( (\DM)^2(t^2-n^2)+n^2 \big)},
\label{T-slow-def}
\\
\hat\TT_n^\mathrm{pass}
\=
-
\bar\TT_n^\mathrm{pass}
\sin 2 ( \phi_{\mathrm{s}} t+\tphi ).
\label{T-fast-def}
\end{gather}
%
Formula 
\eqref{T-pass-def}
yields the asymptotic decoupling of {thermal fundamental solution} 
$\TT_n$ as the sum of the slow 
$\bar\TT_n^\mathrm{pass}$
and the fast $\hat\TT_n^\mathrm{pass}$
motions, which are considered as continuum quantities.
Thus, Eqs.~\eqref{T-pass-def},
\eqref{T-slow-def},
\eqref{T-fast-def},
provide a natural continuum description of the
thermal motions in the system under consideration.

\begin{remark}
Formulae \eqref{T-slow-def},
\eqref{T-fast-def}
for all $n\neq0$ transform 
as $m\to1$ into the expressions for the slow
and the fast motions, respectively, in a uniform chain. The last expressions are 
obtained in \cite{Gavrilov2022ijhmt}
using the asymptotic estimation of an integral representation for the exact solution 
in the explicit form \eqref{Sro-bessel} at a
moving point of observation.
\end{remark}

\begin{remark}  
In study \cite{Kuzkin-Krivtsov-accepted} an alternative procedure of
uncoupling of the slow and the fast motions was suggested for the case of an
arbitrary uniform scalar harmonic lattice. In the case of a uniform chain, 
our approach leads to the same
expression for the slow motion, and different expression for the fast motion.
Apparently, the result of 
\cite{Kuzkin-Krivtsov-accepted} related with the fast motion can be obtained 
by calculation of the discrete convolution of the fundamental solution
$\TT_n$ in the form of Eq.~\eqref{T-pass-def} with a slowly spatially-varying
discrete function describing initial conditions for the particle velocity. 
Nevertheless, this issue requires an additional
investigation.
\end{remark}


\begin{remark}
In the particular case $m=1$ formula \eqref{T-slow-def} coincides
for $n\neq0$ 
with the solution of the ballistic
heat equation introduced in \cite{krivtsov2015heat,krivtsov-da70}, which
provides continuum description of heat transport in a uniform chain. The
corresponding formula in this particular case also can be found in recent
study \cite{Allen2022}, where an alternative approach to continuum description of heat
transport in a uniform chain is suggested.
\end{remark}

\begin{remark}  
{Since the term of order $t^{-1/2}$ for the expansion of $\VF_n^\mathrm{pass}$
\eqref{I_1+I_2} and the approximate solution $\VA_n^\mathrm{pass}$ 
\eqref{v-decay}
are zeros for $n=0$,
strictly speaking, we should use 
Eq.~\eqref{rubin-f} instead. This yields:}
\begin{equation}
\bar\TT_0=\frac{m^2}{4\pi(\DM)^4\, t^{3}}
+ o(t^{-3}).
\end{equation}
\end{remark}

In the case $m<1$ due to 
\eqref{v-pass-stop}
one has
\begin{equation}
(\VA_n)^2
=
(\VA^\mathrm{pass}_n)^2+(\VA^\mathrm{stop}_n)^2
+2\VA^\mathrm{pass}_n\VA^\mathrm{stop}_n,
\end{equation}
and, therefore,
\begin{equation}
\TT_n\simeq
\TT_n^\mathrm{pass}
+
\TT_n^\mathrm{stop}
+
\TT_n^\mathrm{cross},
\end{equation}
where the first term $\TT_n^\mathrm{pass}$ is defined by 
Eq.~\eqref{T-pass-def}. Calculating the second term $\TT_n^\mathrm{stop}$
using \eqref{v-loc}, one gets
\begin{equation}
(\VA^\mathrm{stop}_n)^2
=
\frac{2(\DM)^2m^{2(|n|-1)}}{(2-m)^{2(|n|+1)}}
(1+\cos 2\Omega_0 t).
\label{sq-v-st-band}
\end{equation}
Hence,
\begin{gather}
\TT_n^\mathrm{stop}=
\bar\TT_n^\mathrm{stop}
+
\hat\TT_n^\mathrm{stop},
\\
\bar\TT_n^\mathrm{stop}
\=
\frac{4m_n(\DM)^2m^{2|n|-1}}{(2-m)^{2(|n|+1)}},
\label{T-skow-stop-def}
\\
\hat\TT_n^\mathrm{stop}
\=
\bar\TT_n^\mathrm{stop}
\cos 2\Omega_0t.
\label{T-fast-slow-def}
\end{gather}

\begin{remark}  
Expression 
\eqref{T-skow-stop-def} for 
the slow motion $\bar\TT_n^\mathrm{stop}$ can be continualized by a natural
way in the range $|n|\geq1$, where $m_n=1$.
\end{remark}

Finally, for the cross product term $\TT_n^\mathrm{cross}$ one gets
\begin{equation}
\TT_n^\mathrm{cross}\=
4mm_n\VA_n^\mathrm{pass}\VA_n^\mathrm{stop}
\end{equation}
and, therefore,
\begin{multline}
\TT^\mathrm{cross}_n=
\frac{8m_n|n|(\DM)(-1)^{|n|+1} m^{|n|}
\,\cos\left( \phi_{\mathrm{s}} t+\tphi-\frac{\pi}{4} \right)
\,\cos\Omega_0t
\,\Hf}
{\pi^{1/2}(t^2-n^2)^{1/4}((\DM)^2(t^2-n^2)+n^2)^{1/2}(2-m)^{|n|+1}}
\\
=
\frac{4m_n|n|(\DM)(-1)^{|n|+1} m^{|n|}
\,\Hf}
{\pi^{1/2}(t^2-n^2)^{1/4}((\DM)^2(t^2-n^2)+n^2)^{1/2}(2-m)^{|n|+1}}
\qquad
\\
\qquad{\times
\left(
\cos\left( \left (\phi_{\mathrm{s}}-\Omega_0 \right) t+\tphi-\frac{\pi}{4} \right ) 
+\cos\left( \left(\phi_{\mathrm{s}} + \Omega_0\right) t+\tphi-\frac{\pi}{4} \right )  
\right)}.
\label{sq-v-cross}
\end{multline}
Provided that 
\begin{equation}
\phi_{\mathrm{s}}\pm\Omega_0\not\simeq 0, 
\label{cross-simeq}
\end{equation}
quantity 
$\TT_n^\mathrm{cross}$ oscillates in time, and it should be
considered as a component of a fast motion. 
According to 
\eqref{phi_ast-expr} $-2<\phi_\mathrm{s}<0$ for all $w$ satisfying 
inequality \eqref{w-0-1}. On the other hand $\Omega_0\in\mathbb S$. 
Thus, the only case when the cross-term 
\eqref{cross-simeq} 
should be taken into account in the slow motion associated with thermal
transport is 
\begin{equation}
\left\{
\begin{aligned} 
&w\to+0,
\\
&m\to1-0;
\end{aligned}
\right.
\quad \Longleftrightarrow\quad 
\left\{
\begin{aligned} 
&\phi_\mathrm{s}\simeq-2+0,
\\
&\Omega_0\to2+0;
\end{aligned}
\right.
\end{equation}
which is beyond the scope of the current
paper (see Sect.~\ref{sect-disc}).

Hence, we obtain that in the case $0<m<1$ the slow motion
$\bar\TT_n$ can be approximately described by
formula 
\begin{equation}
\bar\TT_n\simeq
\bar\TT_n^\mathrm{pass}
+
\bar\TT_n^\mathrm{stop};
\end{equation}
where 
$\bar\TT_n^\mathrm{pass}$ 
and 
$\bar\TT_n^\mathrm{stop}$ 
are given by Eqs.~\eqref{T-slow-def}
and \eqref{T-skow-stop-def}, respectively. The first term in the right-hand
side of the last formula is associated with the thermal transport
outside from the source, whereas the static second term is associated with the
thermal energy trapped near the source. The fast motion is 
\begin{equation}
\hat\TT_n\simeq
\hat\TT_n^\mathrm{pass}
+
\hat\TT_n^\mathrm{stop}
+
\TT_n^\mathrm{cross},
\end{equation}
wherein the terms in the right-hand side are defined by Eqs.~\eqref{T-fast-def},
\eqref{T-fast-slow-def},
\eqref{sq-v-cross}.

\begin{remark}
Consider the following inequality: $\bar\TT^\mathrm{stop}_0/\bar\TT_1^\mathrm{stop}>1$.
Using Eq.~\eqref{T-fast-slow-def} this inequality can be equivalently
rewritten as $m^2-5m+4>0$, which is true for $0<m<1$. Thus, for $t\to\infty$
we always observe
``a hot point'' at $n=0$ in the case of a light defect.
\end{remark}

\begin{figure}[htb]  
\centering\includegraphics[width=0.9\textwidth]{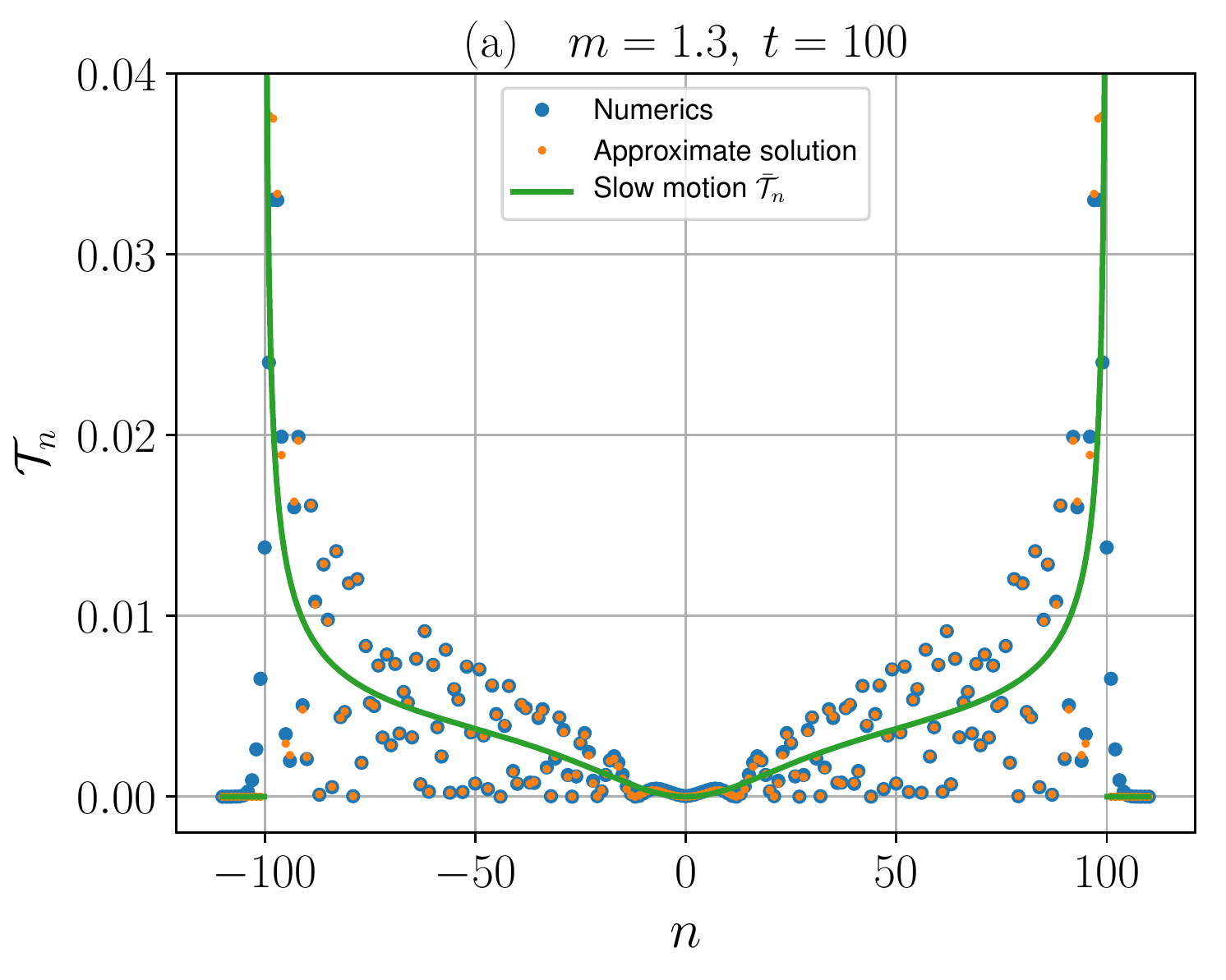}
\centering\includegraphics[width=0.9\textwidth]{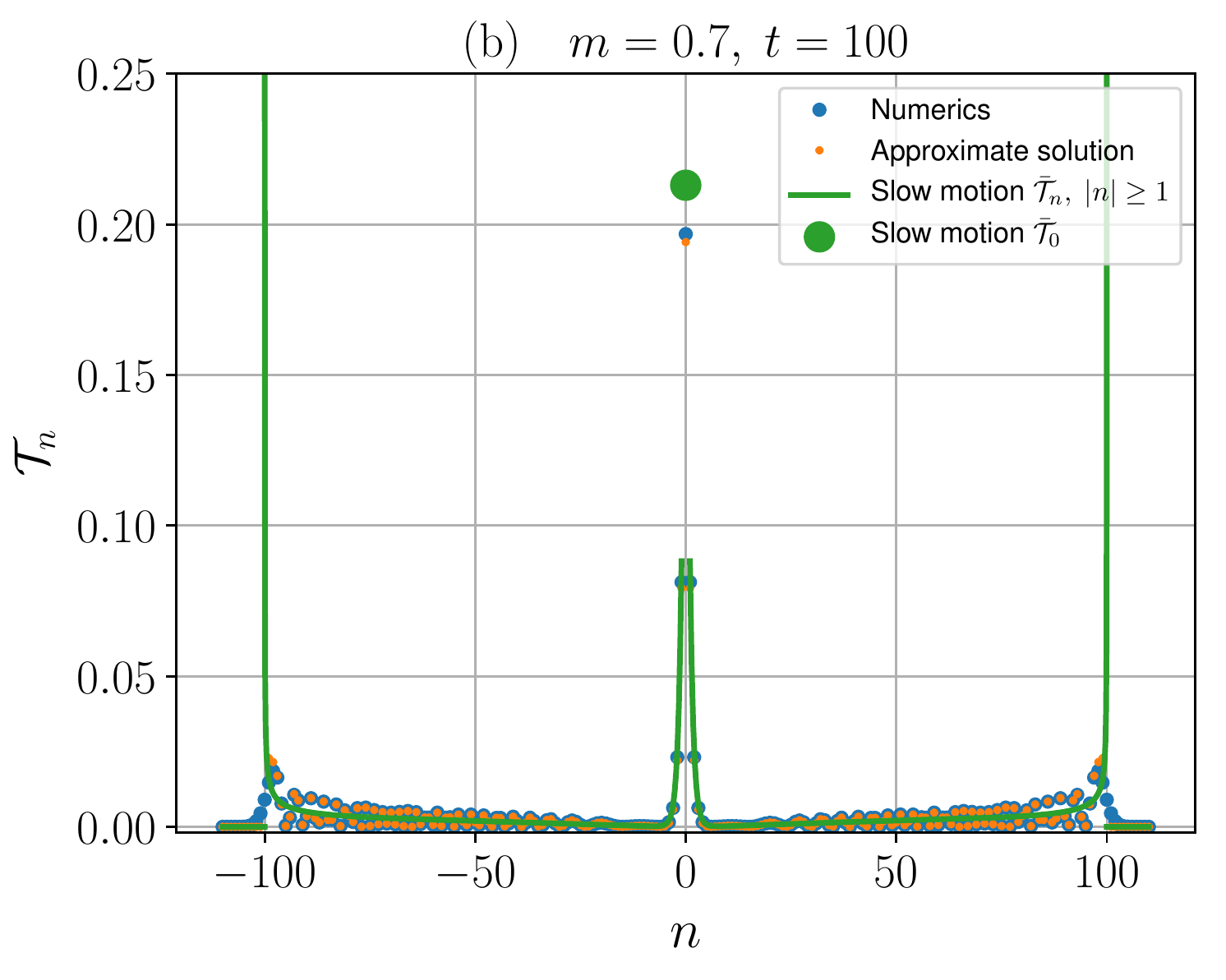}
\caption{Comparing the approximate solution for $\TT_n$
in the form of 
Eq.~\eqref{TT-app},
the corresponding solution 
\eqref{thermal-fs}
for the kinetic temperatures wherein $\VF_n$ are found numerically,
and the slow motion $\bar\TT_n$.
(a) The case of
a heavy defect, (b) the case of a light defect (here $\bar\TT_n$ is represented as a
continuum quantity for $|n|\geq1$)}
\label{T.pdf}
\end{figure}

In Fig.~\ref{T.pdf} we compare the approximate solution for $\TT_n$
in the form of 
Eq.~\eqref{TT-app},
the corresponding solution 
\eqref{thermal-fs}
for the kinetic temperatures wherein $\VF_n$ are found numerically,
and the slow motion $\bar\TT_n$ for the cases of a
heavy defect and a light one. One can see that the slow motion looks like
a spatial average for $\TT_n$ (everywhere excepting a neighbourhood of the
wave-front $|n|=t$ and a neighbourhood of a light defect).

\begin{figure}[htb]  
\centering\includegraphics[width=0.9\textwidth]{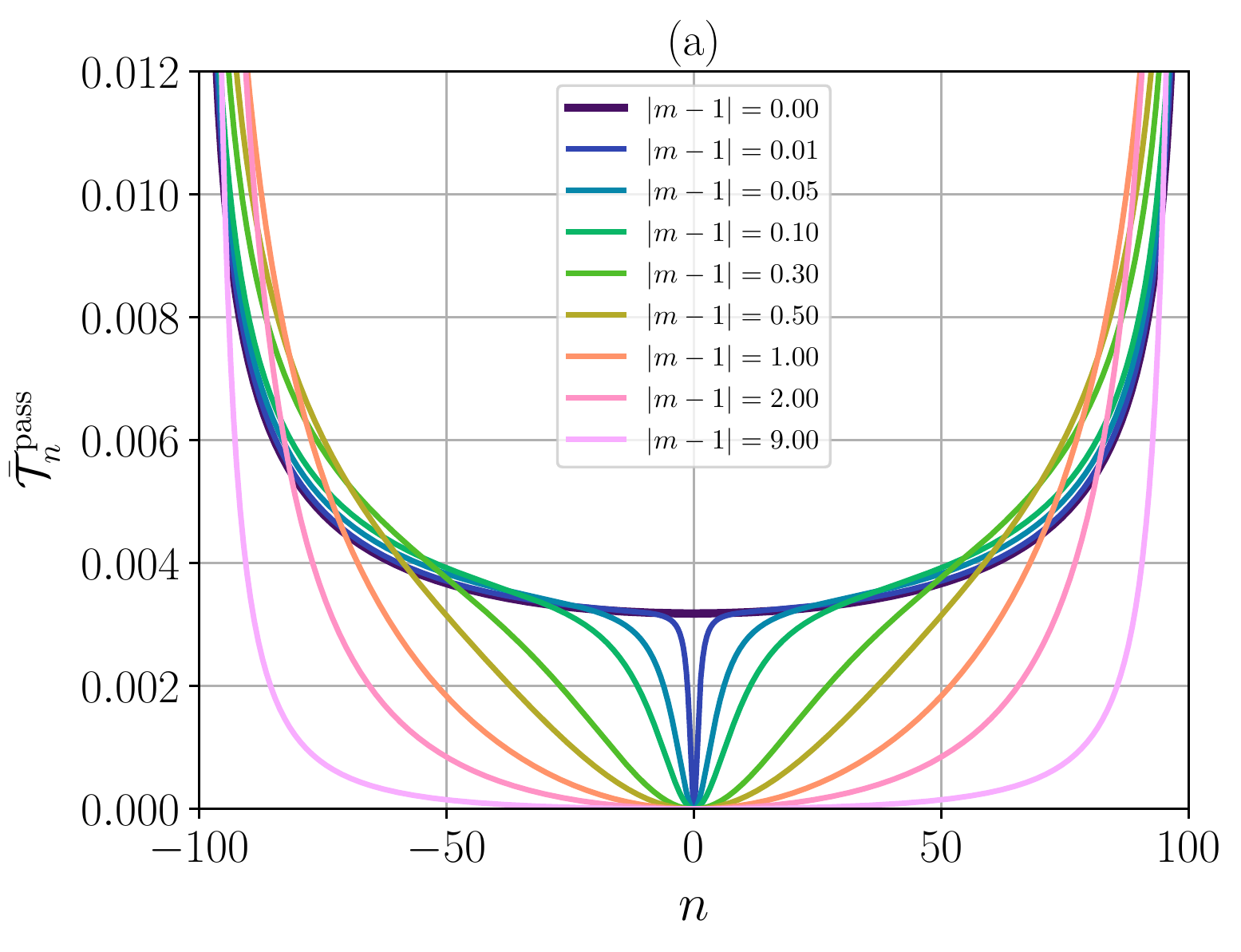}
\centering\includegraphics[width=0.9\textwidth]{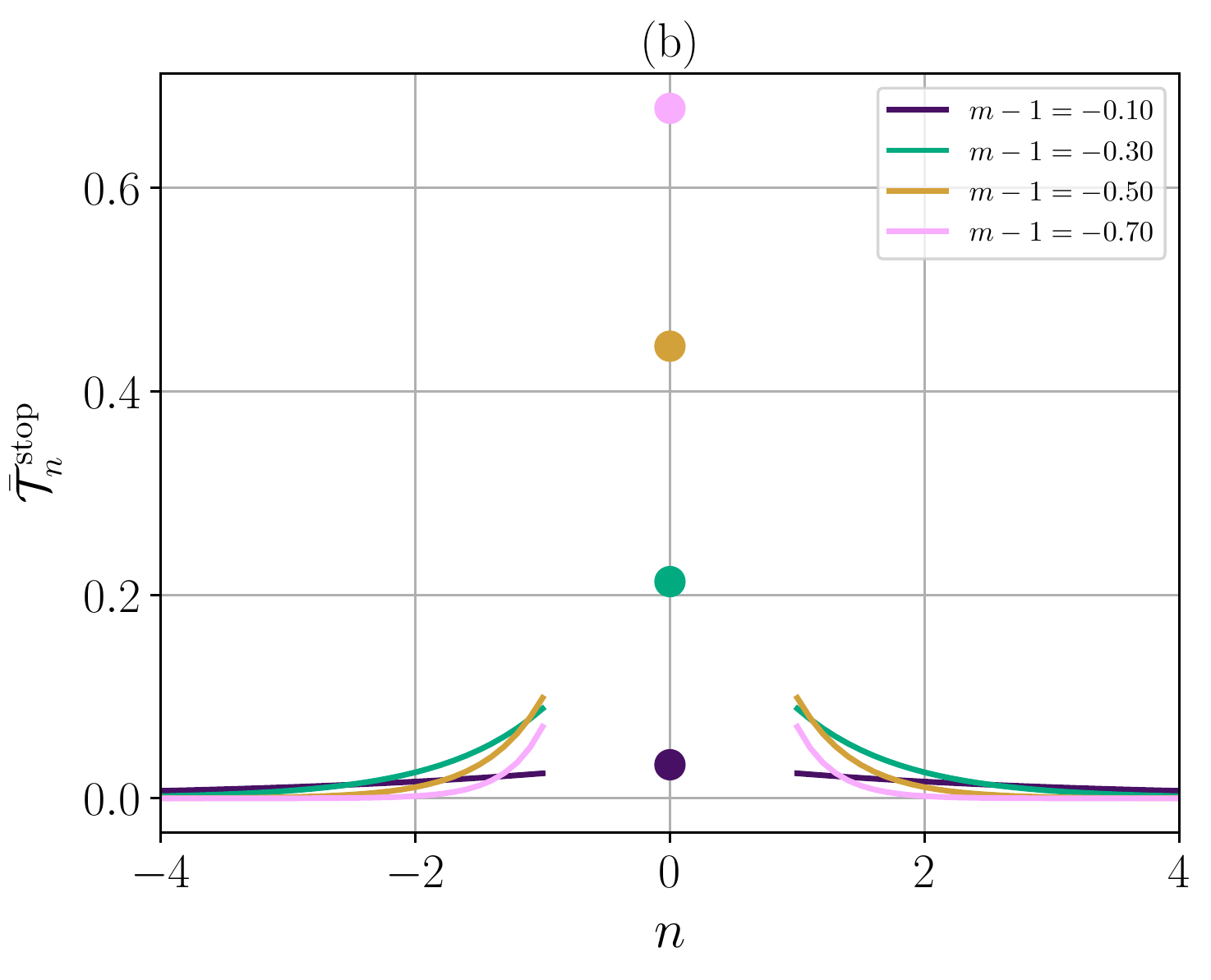}
\caption{The components of the slow thermal motion $\bar{\mathscr T}_n$ 
for various mass of isotopic defect $m$.
(a) $\bar\TT_n^\mathrm{pass}$ given by 
Eq.~\eqref{T-slow-def},
(b) $\bar\TT_n^\mathrm{stop}$ given by 
Eq.~\eqref{T-skow-stop-def}}
\label{T-slow.pdf}
\end{figure}

In Fig.~\ref{T-slow.pdf} we demonstrate the plot for the components 
$\bar\TT_n^\mathrm{pass}$ and 
$\bar\TT_n^\mathrm{stop}$
of the
slow thermal motion $\bar{\mathscr T}_n$ versus $n$ for various values of
$m$. 
One can observe that the greater the quantity $|\DM|$, the wider the
anti-localization zone near the defect, see Fig.~\ref{T-slow.pdf}(a). 
On the other hand, the localization
effect is most pronounced for $m\to+0$.

\afterpage{\clearpage}

\section{Conclusion}
\label{sec-conc}
In the paper we have applied the asymptotic technique based on the method of
stationary phase and have obtained the approximate large-time description of the thermal 
motion\footnote{Thermal motion corresponds to the propagation
of the kinetic temperature.} 
caused by a source on an isotopic defect in a 1D harmonic crystal. 
The most
important new results of the paper are formulae
\eqref{T-pass-def},
\eqref{T-slow-def}, 
\eqref{T-fast-def},
which provide continualization 
and asymptotic uncoupling of the propagating outside the defect
component of the thermal motion into the superposition of the slow and fast motions. The slow
motion is related with ballistic heat transport, whereas the fast motion
is energy oscillation related  with the transformation of the kinetic energy into
the potential one and in the opposite direction. 

In the case of a heavy defect the thermal motion can be described by
the propagating part only, and both slow and fast components are zero at the
defect, see Fig.~\ref{T.pdf}(a).
We suggest calling this phenomenon ``anti-localization''. The physical
meaning of the anti-localization is a tendency for the unsteady propagating
wave-field to avoid a neighbourhood of a defect. 
In the paper we first time demonstrate how the anti-localization influences
on the propagating wave-field. 
Namely, one can observe that the greater the quantity  $|\DM|$ (the dimensionless 
difference between
the mass of the defect and mass of any other particle), the wider the
anti-localization zone near the defect, and more energy concentrates closer to
the leading wave-front $|n|=t$, see Fig.~\ref{T-slow.pdf}(a).
For the best of our knowledge,  the anti-localization was never treated as a general
phenomenon. But its existence under certain, still unknown, conditions is in an excellent agreement both with our
results and observations in studies
\cite{Rubin_1963,Takizawa1968,kaplunov1986torsional,Mueller1962,Kashiwamura1962,Gendelman2021}
concerning the same or different dispersive systems with
inclusions or defects.

In the case of a light defect the slow motion is the superposition of anti-localized
propagating and
localized components. Localized component is a non-vanishing undamped
oscillation, which traps some portion of the full energy and conserves it
forever.  The most complete description of the localized component is
obtained by Rubin in \cite{Rubin_1963} and has been discussed in the current
paper by means of a different approach.

To obtain the vanishing propagating
component of the fast and slow motion we have estimated the exact solution in the
integral form at a moving point of observation. 
This approach allows one to describe running waves, wave-fronts, and to
describe the wave-field as a whole. Thus, the obtained solution is valid in
a wide range of a
spatial co-ordinate (i.e., a particle number), everywhere excepting a neighbourhood
of the leading wavefront.
In previous studies only particular results were obtained, which characterize
the solution at some points and for particular values for the mass of the
defect
(see Remark~\ref{remark-rubin}). Formula 
\eqref{v-onfront} describing the fundamental solution for particle velocity at
the leading wavefront, for the best of our knowledge, is also new. All our 
results have been verified numerically, and a good agreement for the case
$m\not\simeq1$ has been demonstrated. On the other hand, all our results (as well
as the previous ones) become 
practically inapplicable for small $n$ at the case $m\simeq1$, $m\neq1$ (see
Sect.~\ref{sect-disc}). To analyse this case it is necessary to construct a uniform 
asymptotics,
which takes into account the singularities collocations.

Finally, let us discuss how the results of the paper can be generalized, and
how one can 
use them in physical applications. In
our opinion, the same approach can be applied to various more complicated
non-uniform systems (discrete or continuum ones)
with a single point inclusion or a defect.
The derivation of the
expression for the slow motion describing heat transfer 
in a polyatomic harmonic lattice with an isotopic defect
(in particular, in a graphene lattice) could be
a possible direction of the future work. 
The method also can be applied to
the problem where the source is
located outside the defect particle.
Besides, we expect that the solution describing the slow motion for the problem
concerning ballistic heat transport in the same system with suddenly applied
point heat source of a constant intensity
\cite{Gavrilov2022cmat,Gavrilov2019cmat,Gavrilov2020cmat} can be obtained by
time integration of the slow motion found in this paper. 
In our opinion, the results of the paper and such a future work 
can allow us to improve the existing theoretical models for ballistic
thermal transport in isotopically modified graphene \cite{Chen2012}.

Another important application of the approach proposed in this paper is
related with the possibility to suggest simple analytical description of
non-stationary Kapitza thermal resistance, basing on the model proposed in
\cite{Paul2020,Gendelman2021}.

%
\section*{Acknowledgements}
The authors are grateful to A.M.~Krivtsov, O.V.~Gendelman, A.~Politi,
Yu.A. Mochalova, V.A.~Kuzkin, A.A.~Sokolov, \framebox{D.A.~Indeitsev,} A.P.~Kiselev, 
S.D.~Liazhkov, D.V.~Korikov, N.G.~Shvarev for useful and stimulating discussions.


\appendix
\section{Non-dimensionalization}
\label{app-nondim}
The equations of motion for the system under consideration are
\begin{gather}
\tilde m_n\frac{\d^2 {\tilde u}_n}{\d\tilde t^2}
-\tilde C(\tilde u_{n+1}-2\tilde u_n+\tilde u_{n-1})=
\delta_n \tilde p\big(\tilde t\big).
\label{dim-chain-eq-basic-al0}
\end{gather}
Here 
$n \in \mathbb{Z}$, $\tilde t$ is the time,
$\tilde u_n(\tilde t)$ is the displacement of the particle with a number
$n$,
$\tilde m_n$ is
the mass of a particle with number $n$:
\begin{gather}
\tilde m_n=\tilde M+\delta_n(\tilde m-\tilde M),
\end{gather}
$\tilde C$ is the bond stiffness. The dimensionless
equations of motion \eqref{chain-eq-basic-al0} can be obtained by introducing 
the following dimensionless quantities:
\begin{equation}
u_n=\frac{\tilde u_n}{\tilde A};\quad
t=\omega \tilde t;\quad
p=\frac {\tilde p}{\tilde C\tilde A},\quad
m=m_0=\frac {\tilde m}{\tilde M}.
\end{equation}
Here $\tilde A$ is the lattice constant (the distance between neighbouring
particles); $\omega\=\sqrt{{\tilde C}/{\tilde M}}$.

\section{The Erd\'elyi lemma}
\label{sec-erdelyi}
\begin{theorem} Let $a>0,\ \alpha\geq1,\ \beta>0$, $f(\Omega)\in C^\infty$,
$f^{(n)}(a)=0\ \forall n.$ 
Then
\begin{gather}
\int_0^a\Omega^{\beta-1}f(\Omega)\,\EXP{\I t\Omega^\alpha}\,\d\Omega\sim
\sum_{k=0}^\infty c_k t^{-\frac{k+\beta}\alpha},\quad t\to\infty;
\\
c_k=\frac{f^{(k)}(0)}{k!\alpha}\,
\Gamma\left(\frac{k+\beta}\alpha\right)\EXP{\frac{\I\pi(k+\beta)}{2\alpha}}.
\end{gather}
\end{theorem}
The proof can be found in \cite{erdelyi1956asymptotic,Fedoryuk1977}.

In Sect.~\ref{sec-non} we sometimes apply Erd\'elyi lemma to integrals,
where $\beta=1$, $0<\alpha<1$. The corresponding asymptotics can be obtained
by taking $\Omega^\alpha$ as the new integration variable, and applying the
Erd\'elyi lemma to the obtained integral.

\section{The trapped energy ratio}
\label{sec-ratio}
According to Eq.~\eqref{E-def} the initial kinetic energy, as well as the total energy
of the chain for all $t$, is $\mathscr E$. 
On the other hand, according to 
\eqref{sol-T-def},
\eqref{Tc-def}
\begin{equation}
\langle K_n\rangle= \frac12{\mathscr E}\TT_n.
\end{equation}
One has
\begin{equation}
\frac{\sum_{n=-\infty}^{\infty}\langle K_n(t)\rangle}
{\mathscr E}
=
\frac{\sum_{n=-\infty}^{\infty}\TT_n(t)}{2}.
\label{ratio-def}
\end{equation}
Considering the energy trapped near the defect as $t\to\infty$, we substitute
into 
Eq.~\eqref{ratio-def} 
$\TT_n=\TT_n^\mathrm{stop}\simeq
2mm_n(\VA_n^\mathrm{stop})^2$:
\begin{equation}
\frac{\sum_{n=-\infty}^{\infty}\langle K_n(t)\rangle}
{\mathscr E}
=m\sum_{n=-\infty}^{\infty}
m_n(\VA_n^\mathrm{stop})^2=m^2(v_0^\mathrm{stop})^2+2m\sum_{n=1}^\infty(\VA_n^\mathrm{stop})^2.
\label{ratio1-def}
\end{equation}
Using Eq.~\eqref{sq-v-st-band}
to calculate the right-hand side of
\eqref{ratio1-def} at $t=\frac{\pi k}{\Omega_0}$, $k\in\mathbb Z$, $k\to\infty$
(when the trapped kinetic energy equals the trapped total energy),
we obtain the ratio $R$ of the trapped total energy to the total energy of the
chain.
Due to 
\eqref{sq-v-st-band} on has
\begin{equation}
(\VA^\mathrm{stop}_n)^2\Big|_{t=\frac{\pi k}{\Omega_0}}
=
\frac{4(1-m)^2m^{2(|n|-1)}}{(2-m)^{2(|n|+1)}}
.
\end{equation}
Calculating the sum of the geometric series in the right-hand side of 
Eq.~\eqref{ratio1-def}, one, finally, gets 
\begin{gather}
R=\frac{4(1-m)^2}{m(2-m)^2}\left(m +\frac{2q}{1-q}\right)=1-\frac{m}{2-m}.
\label{ratio-final}
\end{gather}
Here 
\begin{gather}
q=\frac{m^2}{(2-m)^2}
\end{gather}
is a common ratio for the geometric series.

\begin{figure}[ht]  
\centering\includegraphics[width=0.7\textwidth]{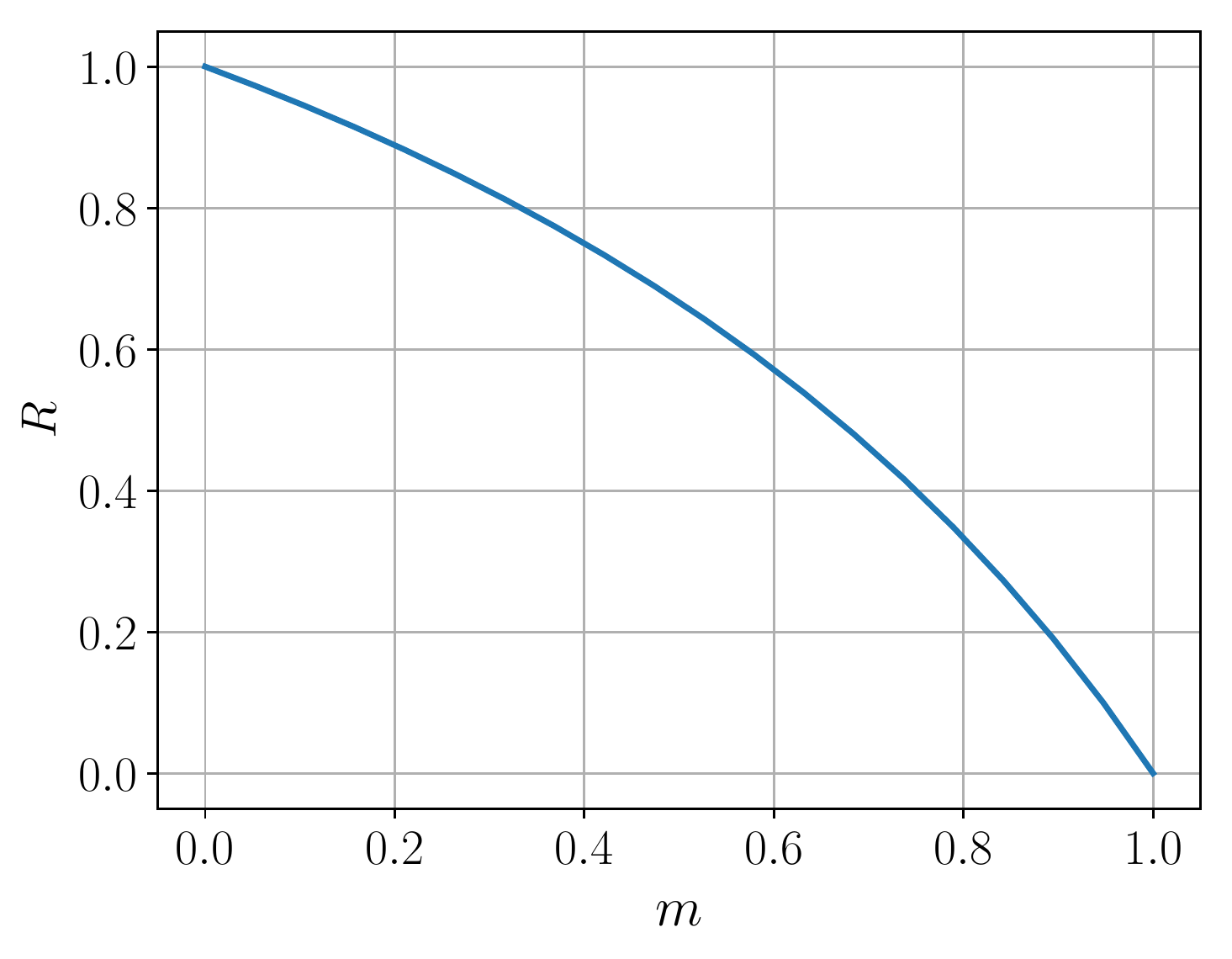}
\caption{The trapped total energy ratio $R$ versus $m$}
\label{trapped.pdf}
\end{figure}

The plot of the trapped total energy $R$ versus $m$ is presented in
Fig.~\ref{trapped.pdf}. For $m\to+0$ all energy is trapped near the defect,
for $m\to 1-0$ all energy is radiated away from the defect.
Formula 
\eqref{ratio-final} was previously obtained in \cite{Rubin_1963}.

\bibliographystyle{spmpsci}
\bibliography{bib/impurity,bib/serge-gost,bib/thermo,bib/math,bib/mode-trans,bib/discrete,bib/graphene,bib/mode}

\end{document}

%% file: def.tex
\def\d{\mathrm d}

\def\PV{\operatorname {PV}}

\def\sign{\operatorname {sign}}

\def\arccosh{\operatorname{arccosh}}

\def\sign{\operatorname{sign}}

\def\Res{\operatorname{Res}}

\def\Re{\operatorname{Re}}
\def\Im{\operatorname{Im}}
\def\I{\mathrm i}

\def\A{\mathscr D}
\def\cc{\mathrm{c.c.}}

\def\Idva{I^\mathrm{pass}}
\renewcommand{\=}{\stackrel{\mbox{\scriptsize def}}{=}}
\def\asst{{\mathrm s}}
\def\pFO{}
\def\tphi{\psi}

\def\EXP#1{\mathrm e^{#1}}
\def\UF{{\mathscr U}}
\def\VF{{\mathscr V}}
\def\TT{\mathscr T}
\def\VA{{v}}
\let\mathscr=\mathcal

\def\Hf{H(t-|n|)}
\def\de{\delta}

\def\DM{m-1}